\documentclass[final, times, 3p, twocolumn, authoryear]{elsarticle}

\usepackage{graphicx}
\usepackage{epsfig}
\usepackage{amssymb}
\usepackage{amsmath}
\usepackage{amsbsy}
\usepackage{bm}
\usepackage[colorlinks=true]{hyperref}
\usepackage{graphicx,subfig,epsfig}
\usepackage{amssymb,amsmath,amsbsy,bm,bbm}
\usepackage{algorithm,algpseudocode,theorem}
\usepackage{natbib}
\usepackage[thinlines]{easytable}

\usepackage{soul}
\usepackage[usenames,dvipsnames]{xcolor}
\definecolor{rd}{rgb}{0.7,0.0,0.0}
\definecolor{bl}{rgb}{0.0,0.0,0.7}

\journal{Journal of Theoretical Biology}

\newcommand\R{\mathbb{R}}

\begin{document}

\begin{frontmatter}

\title{The Impact of Competition Between Cancer Cells and Healthy Cells on Optimal Drug Delivery}

\author[umd]{H. Cho}
\ead{hcho1237@math.umd.edu}
\author[umd,umd2]{D. Levy\corref{cor1}}
\ead{dlevy@math.umd.edu}

\cortext[cor]{Corresponding author.  Email:  dlevy@math.umd.edu.
  Phone:  301-405-5140.  Address: Department of Mathematics,
  University of Maryland, College Park, MD 20742-4015}

\address[umd]{Department of Mathematics, University of Maryland, College Park, MD 20742, USA}
\address[umd2]{Center for Scientific Computation and Mathematical Modeling (CSCAMM), University of Maryland,
  College Park, MD 20742, USA}


\begin{abstract} 
Cell competition is recognized to be instrumental to the dynamics and structure of the tumor-host interface 
in invasive cancers. 
In mild competition scenarios, the healthy tissue and cancer cells can
coexist.
When the competition is aggressive, competitive cells, the so called 
super-competitors, expand by killing other cells.
Novel cytotoxic drugs and molecularly targeted drugs are commonly
administered as part of cancer therapy.
Both types of drugs are susceptible to various mechanisms of drug
resistance, obstructing or preventing a successful outcome.
In this paper, we develop a cancer growth model that accounts for the 
competition between cancer cells and healthy cells.  The model
incorporates resistance to both cytotoxic and targeted drugs.
In both cases, the level of drug resistance is assumed to be a
continuous variable ranging from fully-sensitive to fully-resistant.
Using our model we demonstrate that 
when the competition is moderate, therapies using 
both drugs are more effective compared with
single drug therapies. 
However, 
when cancer cells are highly competitive, 
targeted drugs become more effective. In this case,
therapies that are  initiated with a targeted drug and 
are exposed to it for a sufficiently long time are shown to 
have better outcomes.  
The results of the study stress the importance of adjusting
the therapy to the pre-treatment resistance levels.  We conclude with
a study of the spatiotemporal propagation of drug resistance in a
competitive setting, verifying that the same conclusions hold in the
spatially heterogeneous case.
\end{abstract}

\begin{keyword}
Tumor growth \sep Cell competition \sep Drug resistance \sep Chemotherapy 
\sep Targeted drugs 
\end{keyword}

\end{frontmatter}

\section{Introduction}\label{sec:intro}

\begin{figure}[!htb]
    \centerline{     
   \includegraphics[width=6cm]{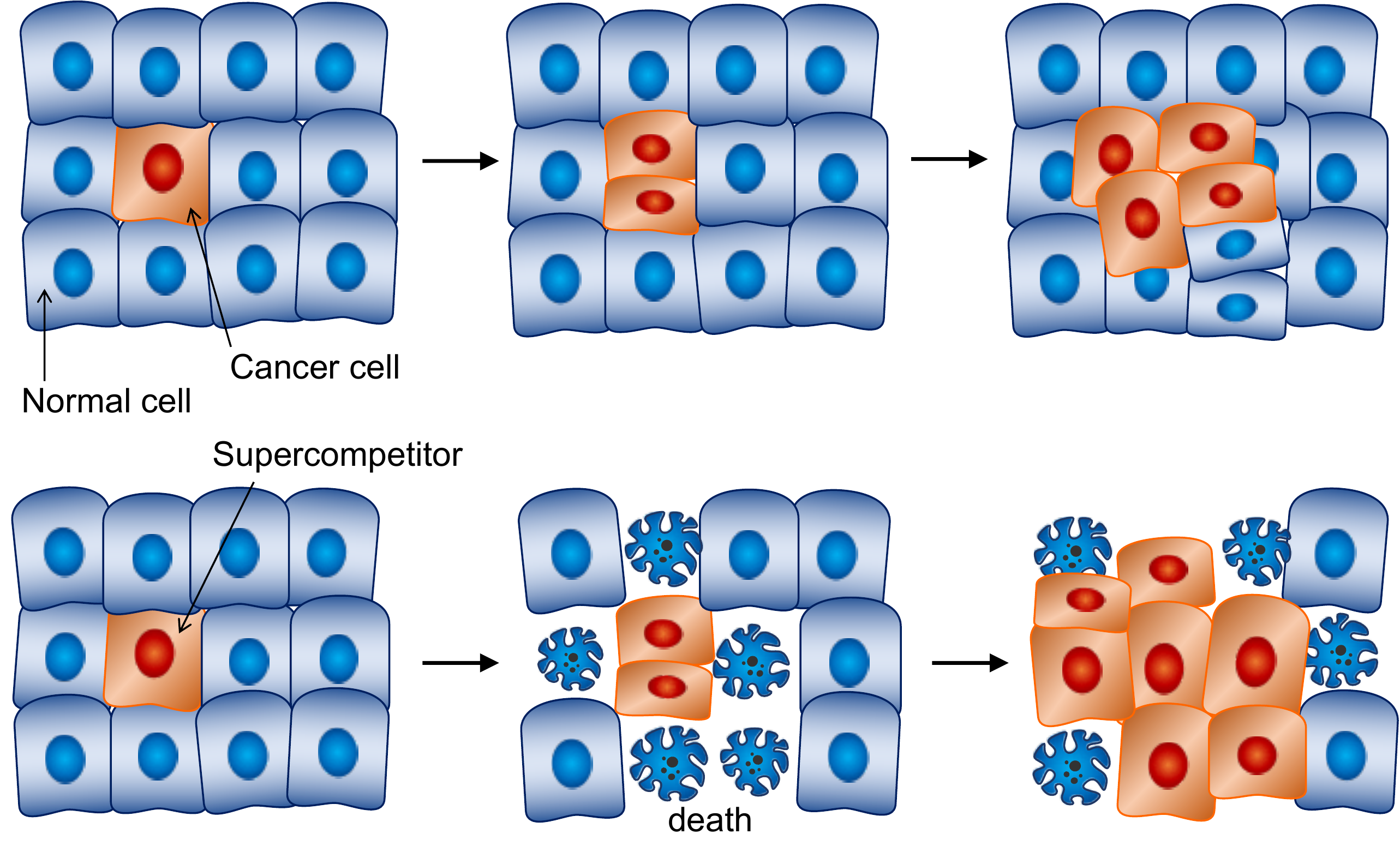} 
    }
   \caption{Competition between 
   healthy cells and cancer cells.
   Often cancer cells and normal cells coexist and 
   the expansion is restricted by spatial competition (top). 
   However, some cancer 
   cells, the so called super-competitors, 
   expand by killing the healthy cells (bottom).  Diagram
   adapted from \cite{Moreno2008}.
    }
\label{fig:0} 
\end{figure}

Intra-tumor heterogeneity that 
results from both genetic and non-genetic mechanisms has been
receiving increased attention in recent years
\citep{Marusyk2012,Maley2017,Hanahan2011,Merlo2006,Gatenby2008}. 
Due to phenotypic and mutagenic diversity, 
cancer can be thought of as an ecosystem 
formed by coexisting populations expressing
abnormal features and different cell types that are
embedded in a heterogeneous habitat of normal tissue 
\citep{Hillen2014}. 
Accordingly, competition between tumor cells and 
healthy cells in the host tissue may play
a key role in cancer growth \citep{Moreno2008,Gil2016}. 
Although the mechanisms are complex and are not fully understood, 
it is known that cells can discriminate their types 
via short-range interactions that quantify the relative 
expression levels of particular proteins.
Accordingly, cell competition occurs in the process of identifying and 
eliminating the less fit cells.
The fitness-induced process generally
eliminates the defective cells, 
such as Minute gene mutated cells 
in D. melanogaster \citep{Simpson1979,Moreno2002}.
However, certain types of 
cancer cells can signal the death of their surrounding tissue in a way that 
promotes their neoplastic transformation.
After the ``loser'' cells disappear from the tissue, 
the ``winner'' cells not only survive but also proliferate
to fill out the void created by the dying cells \citep{Moreno2008}. 
Figure~\ref{fig:0} illustrates cancer growth  
in two distinct competition scenarios.

To eliminate cancer cells and suppress their malignant growth, 
various treatments are available, including 
surgery, chemotherapy, 
immunotherapy, and radiotherapy. 
In particular, cytotoxic chemotherapy and 
molecular targeted approaches represent two 
modes of cancer treatment \citep{Masui2013}.
Whereas chemotherapy uses highly potent chemicals 
to target dividing cells, targeted drugs act 
on specific molecular targets that are associated with cancer. 
Novel therapies and drug substances 
are constantly being developed \citep{Ribeiro2012}. 

For both chemotherapy and targeted therapy 
drug resistance is the predominant factor limiting 
clinical success \citep{Masui2013,Gottesman2010,Teicher,Housman2014}. 
For instance, resistance to chemotherapy includes extrinsic mechanisms that 
prevent the drug from reaching its target
in an active form due to short serum half-life or rapid 
clearance by the kidneys and liver \citep{Slingerland2012,Burris2011}.
Intrinsic cellular mechanisms involve 
increased efflux or decreased uptake, 
enzymatic modification and inactivation of the drug, and alteration of drug targets within the cell 
\citep{Gottesman2002,Gottesman2002a,Fodal2011}. 
Resistance to targeted drugs also relies on various mechanisms 
including 
cellular responses that maintain the signaling despite 
the effective targeting 
or signaling through alteration of downstream effectors, 
and cell survival pathways by disabling apoptosis \citep{Byers2013,Zhang2012}. 
Drug resistance involves genetic and epigenetic 
alternations that either exist prior to the treatment 
or acquired, often induced by the drugs \citep{Gottesman2010,Teicher}. 
Clinical trials of combinations of cytotoxic and targeted drugs
suggest that the complementing mechanisms can be used for 
developing effective therapies \citep{Dorris2017,Ribeiro2012}.

Due to the complexity of the underlying mechanisms and the
multifactorial pathways of tumor growth and drug resistance, 
various mathematical models have been developed 
to describe and investigate the emergence of cancer and its
evolution. 
Modeling approaches include deterministic models using differential
equations \citep{Birkhead1987,Tredan2007,Anderson1998} 
and stochastic models 
including branching process and 
multiple mutations for studying
multi-drug resistance and optimal control of drug scheduling
\citep{Komarova2006,Michor2006,Kimmel1998}. 
These modeling approaches have provided a framework for 
improving early detection, for quantifying 
intrinsic and acquired resistance cells, 
and for designing therapeutic protocols 
\citep{LaviDoron2012,Michor2006,Foo2014,Roose2007,Swierniak2009}. 

On top of that, various  
mathematical models incorporate tumor heterogeneity and 
competition between distinct cell types 
\citep{Bacevic2017,Piretto2018,Lorz_woSp,Yoon2018}. 
Recent models using ordinary differentiation equations (ODEs)
focus on competition between distinct types of cancer cells 
that are either resistant or sensitive to a single drug
(\cite{Piretto2018,Carrere2017,Yoon2018}).
While ODEs can model the overall size of the population, 
partial differential equations can model spatial 
heterogeneity in either the physical space 
or in the phenotypic space. 
Competition models using reaction-diffusion equations 
date back to \cite{Gatenby1996} 
that describe
the spatial distribution and temporal evolution of an
invasive tumor, accounting for
the density of the normal tissue and the neoplastic cancerous tissue. 
Considering phenotypic heterogeneity, 
\cite{Lorz_woSp} developed a model for the competition between
healthy cells and tumor cells that depends on a 
continuous phenotypic variable of  
cytotoxic drug resistance level. 
All aforementioned models consider 
drug resistance to a single drug.

To study the impact of cell competition and the heterogeneity 
in drug resistance,
we develop a phenotypic structured model 
extending the model proposed in \cite{Lorz_woSp}.
Our model consists of healthy cells and tumor cells 
depending on
a continuous variable that represents the  
level of drug resistance.
The model is aimed at designing effective combination therapies
(see also \cite{Perthame2014,Cho2017a}). 
We study two scenarios:
(i) a mild competition that allows coexistence of distinct cells; and  
(ii) an aggressive competition that
results with the elimination of one population 
(see Figure~\ref{fig:0}).
We examine the tumor response under  
a combination therapy of cytotoxic and targeted drugs,
assuming a continuous level of drug resistance to 
each drug. This distinguishes our work 
from most other competition models that 
only consider resistance to a single drug 
\citep{Carrere2017,Lorz_woSp,Bacevic2017}. 
Our study implies that 
the optimal order between 
the drugs as well as
the duration of therapy, depend on 
the competition parameter and on
the ratio of preexisting resistant cells to each drug. 

The paper is organized as follows. In section~\ref{sec:meth}, 
we introduce the competition model 
between cancer cell and healthy cells 
with a multi-dimensional resistance trait. 
We estimate 
the range of the competition rate that corresponds to the
super-competitive scenario, where only
one population can survive. 
In section~\ref{sec:Num}, 
we numerically study our model, focusing on cancer growth 
and on the emergence of resistance 
under different combination therapies.
Therapies in which one drug is switched for a second drug
are compared to 
single-drug therapies in section~\ref{sec:Num1}, 
particularly when the competition rate is low. 
In section~\ref{sec:Num2}, we study the effect of 
different continuum models in the resistance space and 
numerically 
compute the optimal switching time that minimizes 
the overall number of cancer cells in a given time interval.  
Alternating therapies and combination on-off therapies are 
compared in section~\ref{sec:Num3}. 
The model is extended to space and the proposed therapies are studied
in section~\ref{sec:Num4}. 
Concluding remarks are provided in section~\ref{sec:conclusion}.

\section{A Mathematical Model for the Competition Between Healthy
  Cells and Cancer Cells}\label{sec:meth}

To model the competition between healthy cells and cancer cells and
the emergence of resistance, we consider two populations:
healthy cells $n_h(t,\theta)$ and cancer cells $n_c(t,\theta)$.
Both populations describe the number of cells at time $t$ that have a
resistance phenotype $\theta$. 
The variable $\theta = (\theta_1,\theta_2) \in [0,1]^2$
describes the level of drug resistance to cytotoxic drugs ($\theta_1$)
and to targeted drugs ($\theta_2$).  The value 0 corresponds to full
sensitivity to the drug, and the value 1 corresponds to complete
resistance.
For example, the level of resistance to cytotoxic agents can 
be related to the expression level of a gene or a gene cluster that is
linked to the cellular level of drug resistance and proliferation
potential, such as MDR1, ALDH1, CD44 \citep{Hanahan2011,Amir2013,Medema2013}. 
We model the competition of $n_h(t,\theta)$ and $n_c(t,\theta)$ as
a reaction-diffusion system,
\begin{align} 
	\partial_t  n_h(t,\theta) &= \left[ r_h(\theta) - d_h(\theta) (\rho_h(t)) - \mu_h(\theta) c_1(t) \right] n_h  \nonumber \\  
	 &  - a_{hc}(\theta) \rho_c(t) n_h + \nu_h \Delta_{\theta} n_h 
	  \label{eq:GvrnH0}, \qquad  \\ 
	\partial_t n_c(t,\theta) &= \left[ r_c(\theta) - d_c(\theta) (a_{ch} \rho_h(t) + \rho_c(t)) - \mu_c(\theta) c_1(t)  \right. \nonumber \\ 
	  & \left. - \varphi_c(\theta) c_2(t) \right] n_c - a_{ch}(\theta) \rho_h(t) n_c + \nu_c \Delta_{\theta} n_c \label{eq:GvrnC0}.  
\end{align}
The reaction terms involve proliferation, apoptosis, 
and drug effect. 
The first reaction terms 
with $r(\theta) > 0$ model the consumption 
of the resources depending on the resistance level. 
We assume that the proliferation rates satisfy $\partial_{\theta_i} r_h(\theta) \leq 0$ 
and $\partial_{\theta_i} r_c(\theta) \leq 0$, 
corresponding to the assumption 
that resistant cells devote their 
resources to developing and maintaining the drug resistance 
mechanisms (see the experimental evidence in \cite{Misale2015,Mumenthaler2015,Wosikowski}).

The death terms involve the rate of apoptosis,
$d_h(\theta)>0$ and $d_c(\theta)>0$. 
We consider a logistic growth model with 
$\rho_h(t)$ and $\rho_c(t)$ 
being the total numbers of 
normal cells and cancer cells, computed as 
\begin{align*} 
\rho_h(t) = \int n_h(\theta,t) d\theta, \quad 
\rho_c(t) = \int n_c(\theta,t) d\theta, 
\end{align*} 
and $\rho(t) = \rho_h(t) + \rho_c(t) $. 
The carrying capacity of normal cells with phenotype $\theta$, is
given by $r_h(\theta) /d_h(\theta) $.
Key to the model are the competition terms:
apoptosis due to competition 
occurs with rates $a_{hc}$ and $a_{ch}$ 
with respect to the size of the other cell population. 
This resembles the competitive Lotka--Volterra model, (e.g.,
\cite{Gatenby1996,Murray2002}), 
and has been referred to as \emph{competition rate}
in competition models \citep{Piretto2018}. 

The drug effects 
that represent the death 
of cancer cells due to 
the action of cytotoxic and targeted drugs 
are also included in the growth term.
The time-dependent dosages are denoted by $c_1(t)$ for the cytotoxic drug and
$c_2(t)$ for the targeted drug. 
The healthy cells are affected only by the cytotoxic drug with a drug
uptake function $\mu_h(\theta)$.  Cancer cells respond to both the
cytotoxic drug and to the targeted drug with uptake functions 
 $\mu_c(\theta)$, and $\varphi_c(\theta)$, respectively.
As the resistance level increases,  
the cells become more resilient 
to the drugs. This translates to the modeling assumption 
$\partial_{\theta_i} \mu_h(\theta) \leq 0$, 
$\partial_{\theta_i} \mu_c(\theta) \leq 0$, 
and $\partial_{\theta_i} \varphi_c(\theta) \leq 0$ \citep{Lorz_wSp,Mumenthaler2015}. 

Chemotherapy uses highly potent chemicals 
that kill rapidly dividing cells, thus we take 
$\mu_h(\theta)>0$ and $\mu_c(\theta)>0$, and assume that 
the therapy is more effective with sensitive cells.
On the other hand, targeted therapies 
selectively target these cancer-related genetic lesions. Hence, we let
$\varphi_c(\theta)>0$, and assume that targeted drugs do not
affect healthy cells.

The Laplacian operator 
$\Delta_{\theta} = \sum_{i=1}^n \partial^2/\partial {\theta_i^2}$ 
describes the instability 
in the resistance phenotypic space 
with rates $\nu_h$ and $\nu_c$. 
In addition to genetic mutations, epimutations 
contribute to phenotypic instability:
heritable changes in gene expression that do not alter the DNA 
\citep{Brock2009,Glasspool2006,Gupta2011}. 
Recent experiments demonstrated that such non-genetic instability and phenotypic 
variability allow cancer cells to reversibly transit between different phenotypic states 
\citep{Chang2006,Pisco2013,Sharma2010}.

\subsection{Studying the competition parameter} 

Recent studies 
suggest that cell competition is often critical  
in shaping cancer development \citep{Vivarelli2012, Wagstaff2013}. 
In particular, the fitness-sensing process during competition 
that usually eliminates defective cells, has a distinctive behavior 
in pre-cancerous lesions \citep{Moreno2008}. 
By acquiring a ''super-fit'' status, 
these super-competitors
mutated cells can sense the surrounding wild-type cells as ''less
fit'' and signal the death of surrounding tissue that in turn promotes
their neoplastic transformation
\citep{Gil2016}. 
In our model, the competition parameters $a_{hc}$ and $a_{ch}$ 
describe the aggressiveness of the cells towards cells from the other
populations.

To characterize competition scenarios,
we simplify Eqs.~\eqref{eq:GvrnH0}--\eqref{eq:GvrnC0} 
by excluding the diffusion terms, 
and consider parameters as follows.  We will revisit the
diffusion terms in a later section.
The proliferation rate and death rate are 
assumed to be constant.  Cancer cells proliferate $A$ times faster
than healthy cells, that is, $r_H = r$ and $r_C = A r$ 
with $A \geq 1$.  
The apoptosis rates are taken as $d_H = d_C = d$. 
The cross-competition parameter 
is taken as $a_{hc} = d_h a$ and $a_{ch} = d_c a$. 
We provide analytical results 
assuming that the model has a single trait $\theta$,  
so that ${\rho}_h = n_h$ and ${\rho}_c = n_c$.
This is of value since phenotype-structured models 
are known to asymptotically converge to
a Dirac-delta distribution
at few dominating resistance traits
\citep{Lorz2011,Perthame2008}. 
We consider the following scenarios:

$\bullet$ {\em No treatment.}
When no drug is applied ($c_1 = c_2=0$), 
a nontrivial equilibrium of $\rho_h>0$ and $\rho_c>0$ exists 
if $r - d ( \rho_h + a \rho_c ) = 0$ and 
$Ar - d ( a \rho_h + \rho_c ) = 0$.
This reduces to a condition that the two populations can coexist 
when the competition rate satisfies $a \leq 1/A$. 
Otherwise, when the competition rate is $a > 1/A$, 
one of the cell populations must become extinct,
either $\rho_h>0$, $\rho_c=0$  
or $\rho_h=0$, $\rho_c>0$.
We consider this case as the 
super-competitive scenario. 
Since $A$ represents the ratio of 
over-proliferation of cancer cells 
compared to normal cells, 
for larger values of $A$, it is likely 
to be super-competitive for 
a larger range of the competition rate $a$. 

\begin{figure}[!tb]
    \centerline{     
   \includegraphics[width=8cm]{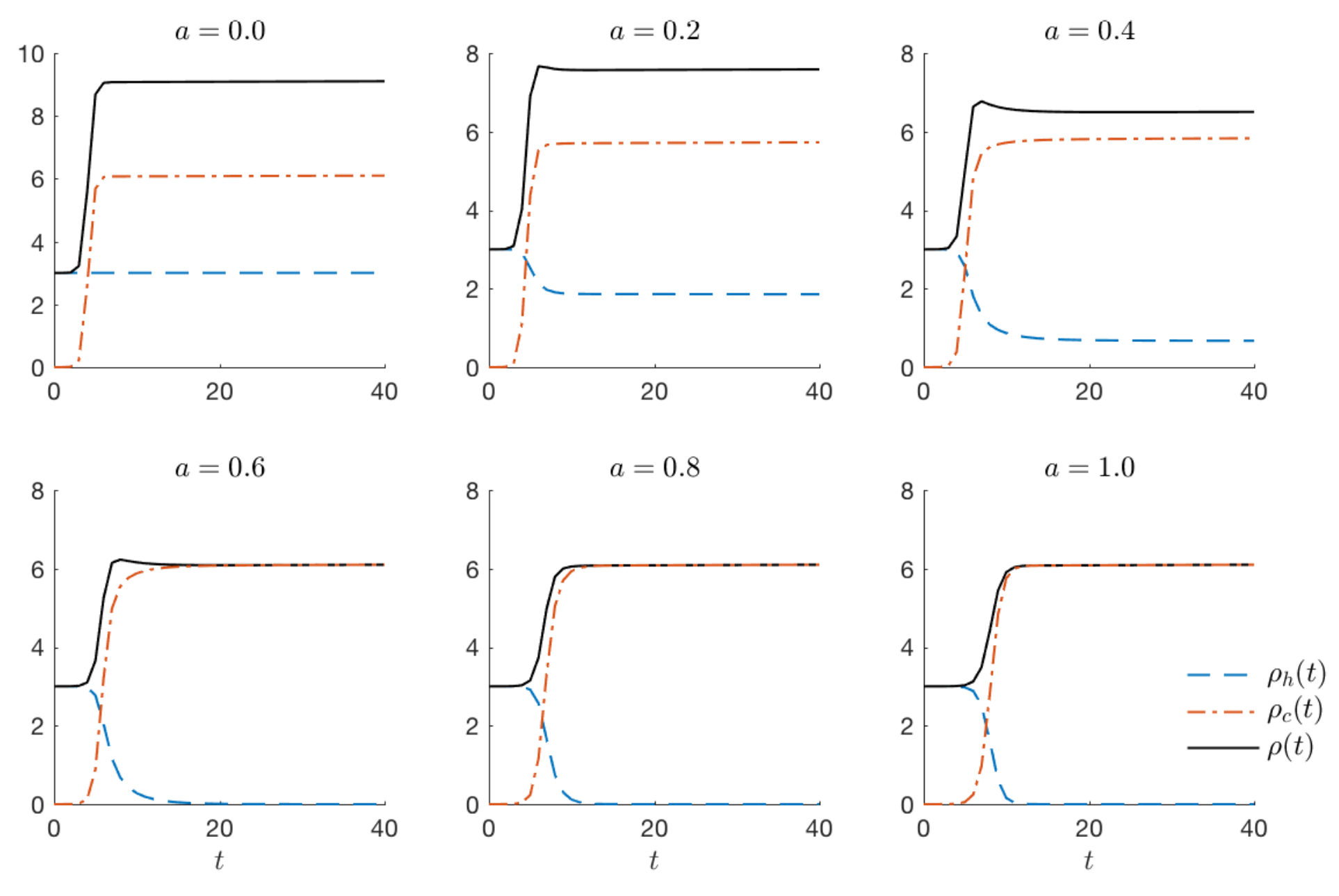} 
    }
   \caption{Comparison of the dynamics of the total
   number of cells $\rho(t)= \rho_h(t)+\rho_c(t)$ 
   for different values of the competition parameter 
   $a = 0,\, 0.2,\, ...,\, 1$.  The cancer cells
   proliferation factor is $A = 2$ and the results are shown up to $t=40$. 
   Healthy cells and cancer cells coexist when $a \leq 1/A = 0.5$, while 
   the cancer cells aggressively overtakes the population 
   when $a > 0.5$. The latter case
   corresponds to the super-competitive model. 
    }
\label{fig:1} 
\end{figure} 

\begin{figure}[!htb]
    \centerline{  \rotatebox{90}{\hspace{0.6cm} \footnotesize (a) No drug }    
       \includegraphics[width=7cm]{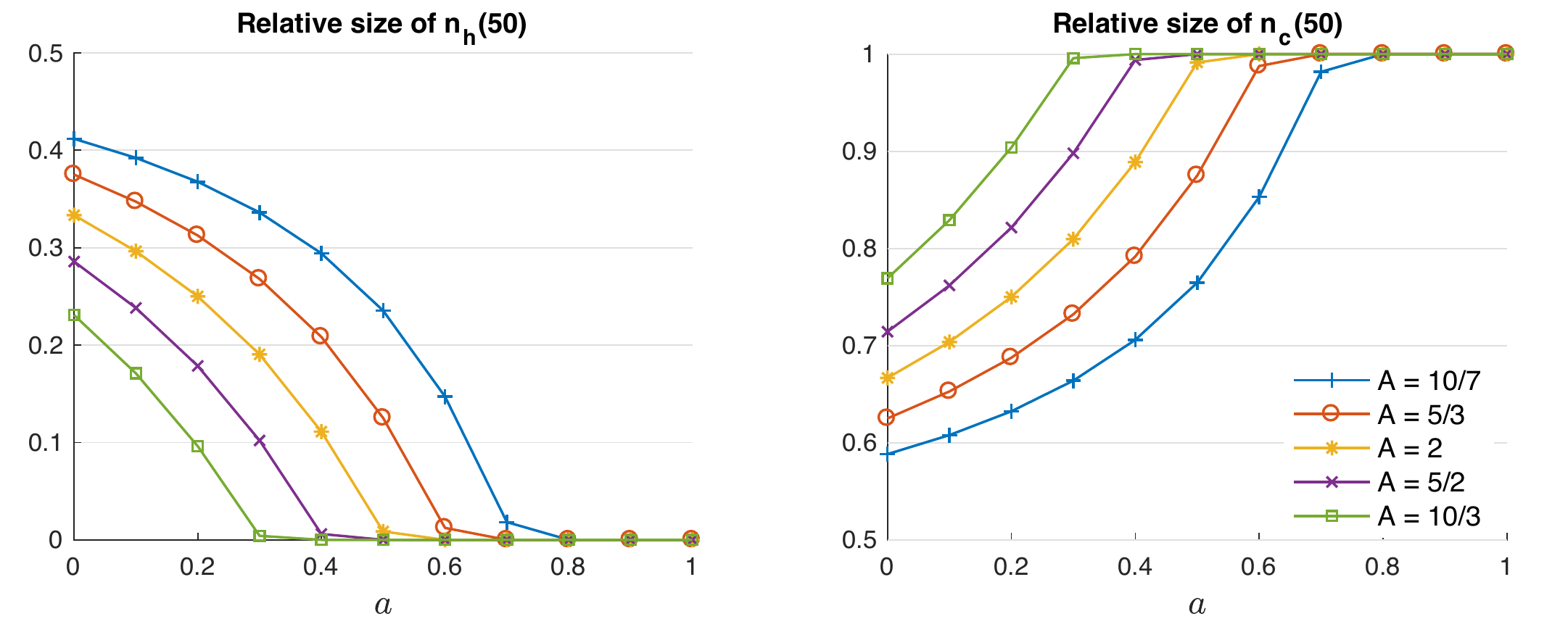} 
    }
    \centerline{  \rotatebox{90}{\hspace{0.6cm} \footnotesize (b) Cytotoxic $c_1$ }    
       \includegraphics[width=7cm]{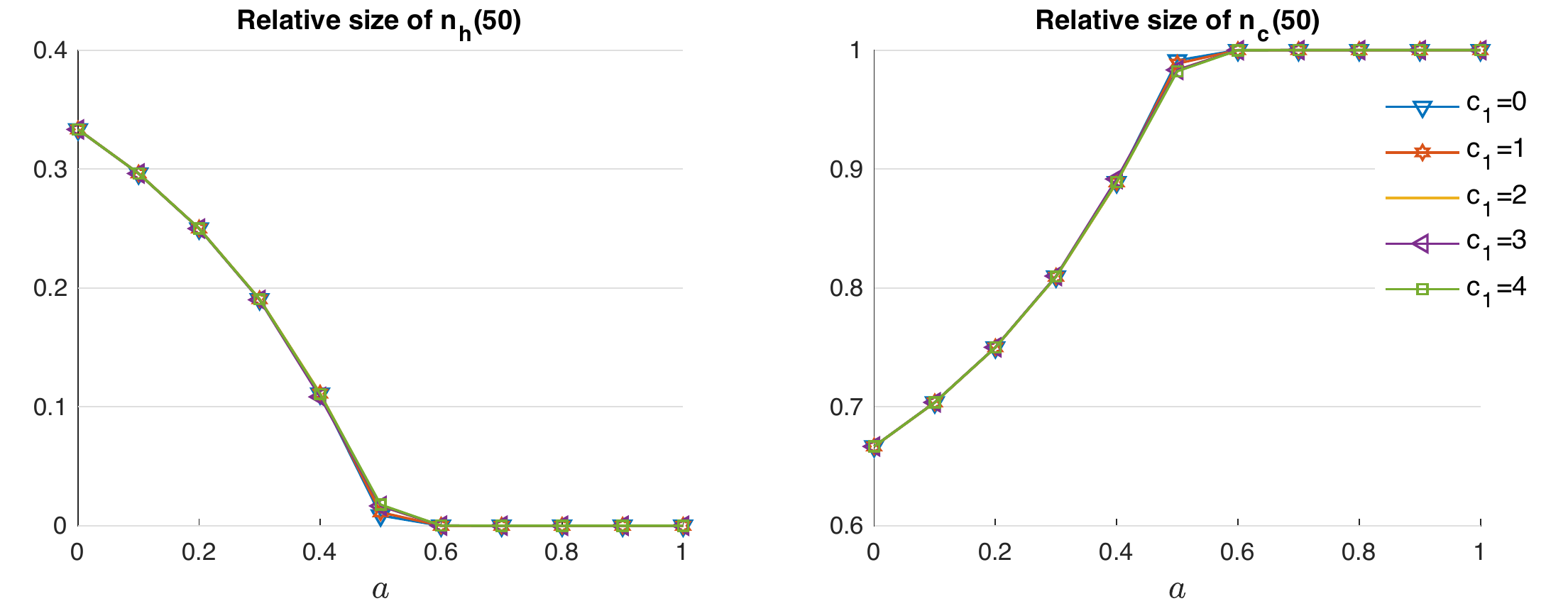} 
    }
    \centerline{  \rotatebox{90}{\hspace{0.6cm} \footnotesize (c) Targeted $c_2$ }    
       \includegraphics[width=7cm]{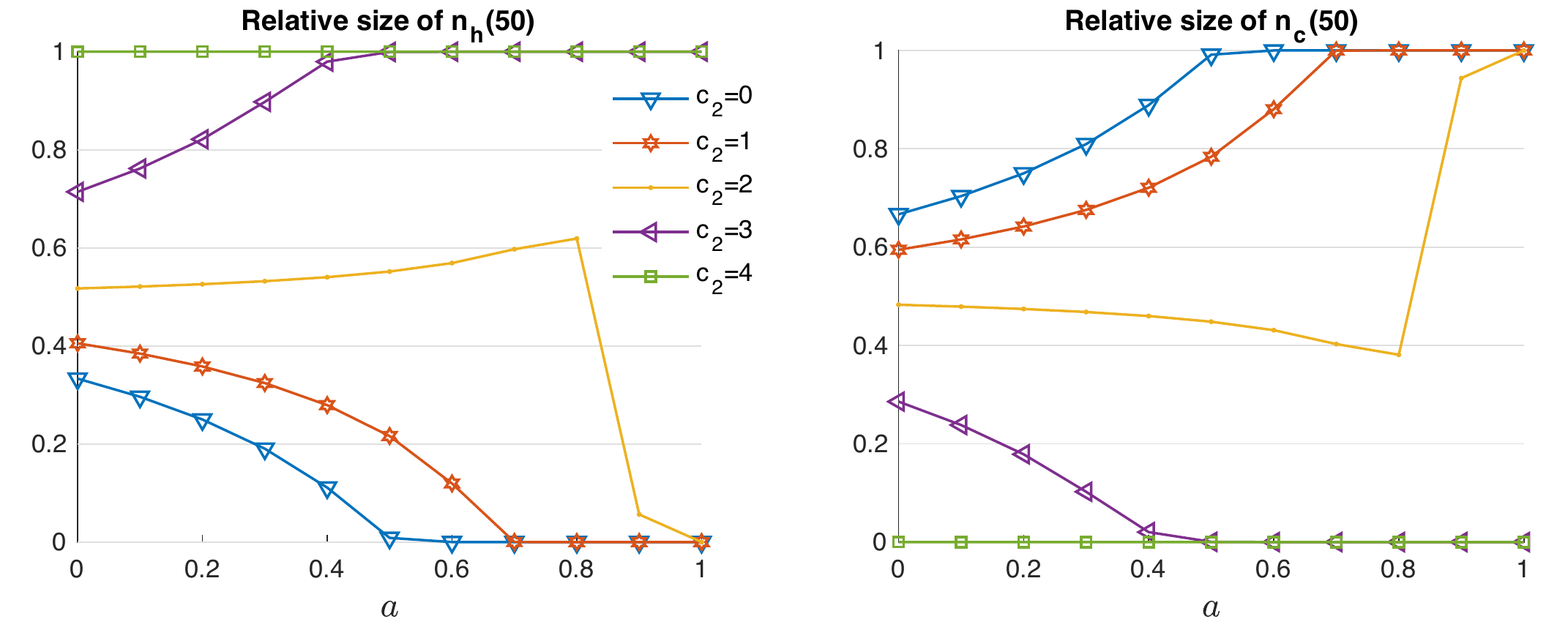} 
    }
   \caption{The relative size of healthy cells $\rho_h(t)/\rho(t)$ (left) 
   and cancer cells $\rho_c(t)/\rho(t)$ (right) at $t=50$ depending on 
   the competition parameter $a$. (a) Without the drug treatment,
   cells coexist when the competition parameter is $a \leq 1/A$. 
   The second and third rows show the results for a fixed $A = 2$. 
   (b) The competition trend does not change 
   when a cytotoxic drug is applied regardless of the dosage.
   (c) The targeted drug has a significant impact on the
   dynamics of coexistence as it only affects the cancer cells.
 }
\label{fig:2} 
\end{figure}

Figure~\ref{fig:1} shows healthy cells, cancer cells, and combined counts,
up to $t = 40$, for $a = 0.0,\,0.2,\,...\,,1.0$.  
We consider $r_h(\theta) = r =  1.5$ and $r_c(\theta) = 2r = 3.0$,
which corresponds to the relative proliferation $A = 2$
Healthy cells and cancer cells 
coexist when $a < 1/2$, but when $a > 1/2$, 
the cancer cells overtakes the population 
and the healthy cells are eliminated. 
This condition is demonstrated again in Figure~\ref{fig:2}(a). The relative numbers of cells 
$\rho_h(t)/\rho(t)$ and $\rho_c(t)/\rho(t)$ at 
$t =50$ are plotted for different values of $A$ and $a$.
The coexistence threshold 
$1/A$ is apparent in the results.

$\bullet$ {\em Weak treatment.}
The condition for coexistence changes when 
the drug is applied with small dosages. 
For the cytotoxic drug, if $c_1 < \min\left( r/\mu_h,\, A r/\mu_c\right)$, 
then coexistence amounts to
$a < (r-\mu_h c_1)/(Ar - \mu_c c_1)$. 
We note that the 
condition is identical to the no treatment case 
if the drug uptake rate of healthy cells versus cancer cells 
is proportional to the proliferation rate, that is, if
$\mu_c = A \mu_h$. 
However, 
under weak targeted therapy, 
$c_2 < r/\varphi_c$, 
coexistence bois down to $a < r /(Ar - A\varphi_c c_2)$ 
so that the cells are likely to coexist for 
larger values of $a$. 

$\bullet$ {\em Strong treatment.}
We assume that the drug 
dosages are sufficiently high 
such that only the most resistant cells can survive, corresponding to
$\theta^* = 1$, where $\theta^* = \arg\max_\theta
(r(\theta)-\mu(\theta)c)$.
With such a dosage $c$, 
the coexistence condition 
is identical to the no treatment case as 
$a \leq 1/A$. 

The results shown in 
Figure~\ref{fig:2}(b,c) are computed with 
$A=2$ and agree with the theoretical thresholds.
While the cytotoxic drug does not change 
the range of $a$ that corresponds to coexistence, the targeted drug 
increases the range of coexistence for weak dosages. 
The threshold of the competition parameter is 
summarized in Table~\ref{Tbl:condA}. 
\begin{table}[!htb] \center 
\begin{tabular}{|c|c|c|} \hline 
No drug, & \multicolumn{2}{|c|}{ Weak drug } \\ \cline{2-3}
& & \\[-1em]
Strong drug & Cytotoxic drug & Targeted drug \\ \hline 
& & \\[-1em]
$a > \dfrac{1}{A}$ &  $a > \dfrac{r-\mu_h c_1}{Ar - \mu_c c_1}$ & $a > \dfrac{r}{Ar - A\varphi_c c_2}$  \\ [-1em]
& &  \\   \hline 
\end{tabular}
\caption{The range of the competition parameter within $0\leq a \leq 1$ 
such that the competition model becomes aggressively competitive, not
allowing for coexistence.
The competition is likely to be aggressive when 
the over-proliferation ratio of the cancer cells $A$ is large. 
We note that the parameter range of the cytotoxic drug reduces to 
$a > {1}/{A}$ for any dosage if $\mu_c = A \mu_h$. 
}
\label{Tbl:condA} 
\end{table}

\subsection{Effectiveness of targeted drugs in the highly competitive case} 

Let us estimate the drug dosage based on 
the dominating trait $\theta^* = \arg\max_{\theta} n_c(\theta)$. 
The instantaneous growth rate at time $t$ is
$$ G_c = r_c(\theta^*) - d_c(\theta^*) (a \rho_h(t) + \rho_c(t)). $$ 
This rate is reduced by
$\mu_c(\theta) c_1(t)$ and $\varphi_c(\theta) c_2(t)$ 
when a drug is administered.
We provide an estimation for the case of
$r_c(\theta^*) = {\eta_c}/({1+\theta^{*2}})$ and 
$\bar{\mu}_c (\theta^*) = \bar{\mu}_c (1-\theta_1^*),$
assuming that an increased level of resistance implies lower levels of
proliferation and drug effect.
For instance, 
when the sensitive cells are dominant such that $\theta_1^* \approx 0$, 
the growth rate without the treatment becomes 
$G_c \approx \eta_c- d(\theta^*) (a \rho_h(t) + \rho_c(t)) $. 
This growth term for the dominant cells is negative when
the initial dosage of cytotoxic drug satisfies
$$ \dfrac{\frac{\eta_c}{1+\theta^{*2}} - {d(\theta^*)}  
(a \rho_h(t) 
+ \rho_c(t))}{\bar{\mu}_c (1-\theta_1^*)} < c_1. $$ 
As $\theta_1^*$ increases, resistant cells arise. 
The cytotoxic drug dosage that is necessary to lower the number of
cancer cells increases and it may reach 
the maximum tolerated dose.
Eventually, when $\theta_1^* \approx 1$, 
the drug effect term, $c_1 \bar{\mu}_c (1-\theta_1^*)$, becomes negligible 
even with high dosages. 
In this case, reducing the growth rate using the competition term 
involving the healthy cells becomes more effective. Thus, when 
$$\dfrac{\eta_c}{1+\theta^{*2}} - c_1 \bar{\mu}_c (1-\theta_1^*) >0, $$ 
targeted therapy is preferable since it can still 
suppress the cancer cells using the competition term  
$d(\theta^*)  a \rho_h(t) $, preserving the normal 
cells $\rho_h(t)$ despite 
the reduced drug effect due to resistance. 
The effect of this term becomes more critical 
for larger values of the competition parameter $a$.
We verify the effectiveness of targeted drugs 
in the highly competitive case in our simulations.

\subsection{Switching drugs as a function of the resistance ratio} 

The choice of drug can be determined by 
considering the ratio of resistance. 
For simplicity we classify the sensitive and resistant cancer 
cells into four groups as 
$\rho_{SS}$, $\rho_{SR}$, 
$\rho_{RS}$, and $\rho_{RR}$, where $S$ and $R$ 
represent being sensitive and resistant to one of the drugs. 
The first index corresponds to the cytotoxic drug and the second index
corresponds to the targeted drug.
The treatment type can be determined by 
comparing $\rho_{SR}$ and $\rho_{RS}$, 
where the drug type with less
resistant cells should be administered first. 
For instance, if the resistant population to the second drug 
is larger, $\rho_{SR} > \rho_{RS}$, 
it is more effective to use the first drug, 
assuming that the effective decay rates due to each drug, 
$\widetilde{c}_1$ and $\widetilde{c}_2$, are identical. 
Since  
$\rho_{SR} (e^{-\widetilde{c}_1 t}-1)  
<  \rho_{RS} (e^{-\widetilde{c}_2 t}-1)$, we have
\begin{align*}
(\rho_{SS}+\rho_{SR}) e^{-\widetilde{c}_1 t}+(\rho_{RS}+\rho_{RR}) <
  \qquad \qquad  \\ (\rho_{SS}+\rho_{RS}) e^{-\widetilde{c}_2 t} +
  (\rho_{SR}+\rho_{RR}),  
\end{align*}
where the left- and right-hand sides represent the 
number of cancer cells after treated for time $t$ 
with the first and the second drug, respectively. 
This result differs from the one obtained in~\cite{Piretto2018},
in which a combination therapy
of cytotoxic chemotherapy and immunotherapy
assuming resistance to only cytotoxic drugs was considered.
When cells that are resistant to cytotoxic drugs are present, 
it was suggested to first apply the cytotoxic drug \citep{Piretto2018}.

\section{Combining cytotoxic and targeted drugs} \label{sec:Num} 

\begin{figure}[!htb]
    \centerline{  
    \includegraphics[width=8.5cm]{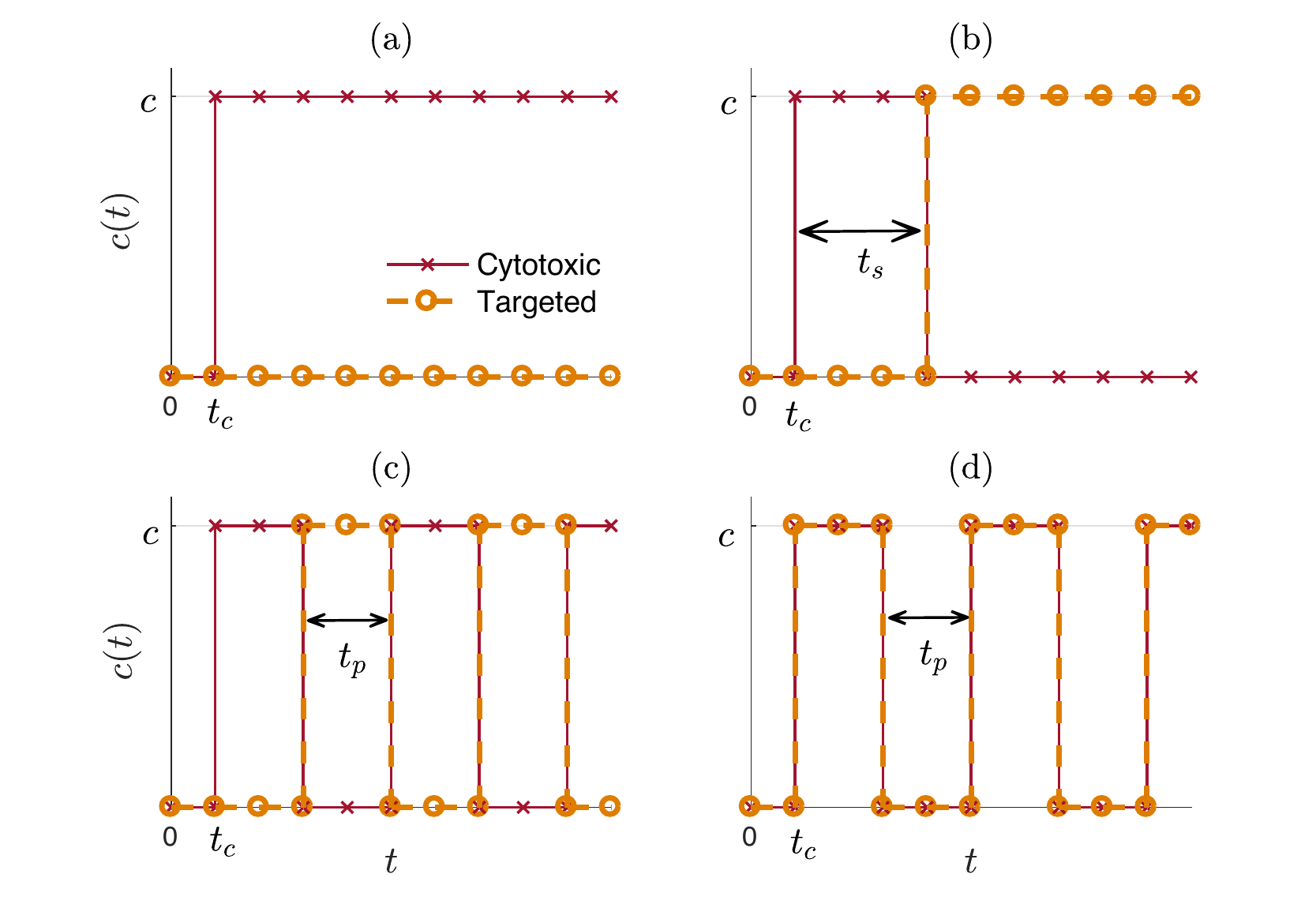} 
    }
   \caption{Drug scheduling considered in our simulations: 
   (a) single cytotoxic drug therapy initiated at $t_c$;
   (b) drug switching therapy such that the drug is switched once after $t_s$;
   (c) alternating therapy with period $t_p$; and (d) on-off combination 
   therapy with period $t_p$. We also test 
   schedules of (a-c) initiated with targeted drugs. 
   }
\label{fig:c(t)} 
\end{figure}

We study the effect of 
different combination therapies with
cytotoxic and targeted drugs. 
The resistance trait becomes $\theta = (\theta_1,\, \theta_2) 
\in [0,\,1]^2$, where $\theta_1$ and $\theta_2$ 
represents resistance to cytotoxic and 
targeted drug, respectively. 
In particular, we consider the model functions as in
\cite{Lorz_woSp}, 
\begin{align}
\label{eq:prolif}
r_h(\theta) &= \frac{\eta_h}{\prod_{i=1}^2 (1+\theta_i^2)},\quad  
r_c(\theta) = \frac{\eta_c}{\prod_{i=1}^2 (1+\theta_i^2)},\,\quad \\ \nonumber
d_h(\theta) &= \frac{d}{\prod_{i=1}^2 (1-0.1\theta_i)},\quad  
d_c(\theta) = \frac{d}{\prod_{i=1}^2 (1-0.3\theta_i)},\, 
\end{align}
where $\eta_h = 1.5$ and  $\eta_c = 3$ are 
the maximum proliferation rates of healthy and cancer cells, respectively,
and $d = 0.5$ is the apoptosis rate. 
The drug uptake functions are taken as 
\begin{align}
\mu_h(\theta) = 0.4(1-\theta_1), \quad  
\mu_c(\theta) = 0.8(1-\theta_1), \nonumber \\
\varphi_c(\theta) = 0.8(1-\theta_2). 
\label{eq:uptake} 
\end{align}
All model functions satisfy the positivity and 
slope assumptions from section~\ref{sec:meth}. 
The drug schedules we consider are shown
in Figure~\ref{fig:c(t)}. 
We consider four different therapies:
(i) a single cytotoxic drug therapy initiated at $t_c$ (a), that is, 
$c_i(t) = c_i \,\bm 1_{t_c \leq t } $;
(ii) a switching therapy such that the drug 
is switched once after $t_s$ (b), 
$c_i(t) = c_i \,\bm 1_{t_c \leq t \leq t_c+t_s} $ and 
$c_j(t) = c_j \,\bm 1_{t_c+t_s \leq t} $;
(iii) an alternating therapy with period $t_p$ (c), 
$c_i(t) = c_i \,\bm 1_{(2n)t_p \leq t-t_c \leq (2n+1)t_p} $ and 
$c_j(t) = c_j \,\bm 1_{(2n-1)t_p \leq t-t_c \leq (2n)t_p}$; 
and (iv) combination on-off therapy with period $t_p$ (d), 
$c_i(t) = c_i \,\bm 1_{(2n)t_p \leq t-t_c \leq (2n+1)t_p} $ and 
$c_j(t) = c_j \,\bm 1_{(2n)t_p \leq t-t_c \leq (2n+1)t_p}$. 

The initial cell populations are set as
\begin{align}
n_h(0,\theta) = \frac{1-w}{C_0} \exp \left[ -\frac{(\theta_1-\mu_1)^2}{\epsilon}-\frac{(\theta_2-\mu_2)^2}{\epsilon}  \right], \nonumber  \\ 
n_c(0,\theta) = \frac{w}{C_0} \exp \left[ -\frac{(\theta_1-\mu_1)^2}{\epsilon}-\frac{(\theta_2-\mu_2)^2}{\epsilon}  \right]. 
\label{eq:IC}
\end{align}
Here, the mean resistance phenotype is 
centered at $\mu_1$ and at $\mu_2$ for the cytotoxic drug and the targeted drug, 
respectively. In addition, $w$ represents the initial proportion 
of the cancer cells in the tissue, and $\epsilon$ controls 
the variance of the preexisting resistance. $C_0$ is a normalizing
constant chosen so that $\rho(0) = 1$. 

\subsection{Single drug and drug switching therapy using cytotoxic and targeted drugs}
\label{sec:Num1}

\begin{figure}[!htb]
    \centerline{      \rotatebox{90}{\hspace{2.4cm} \footnotesize Cytotoxic }   
    \includegraphics[width=7.5cm]{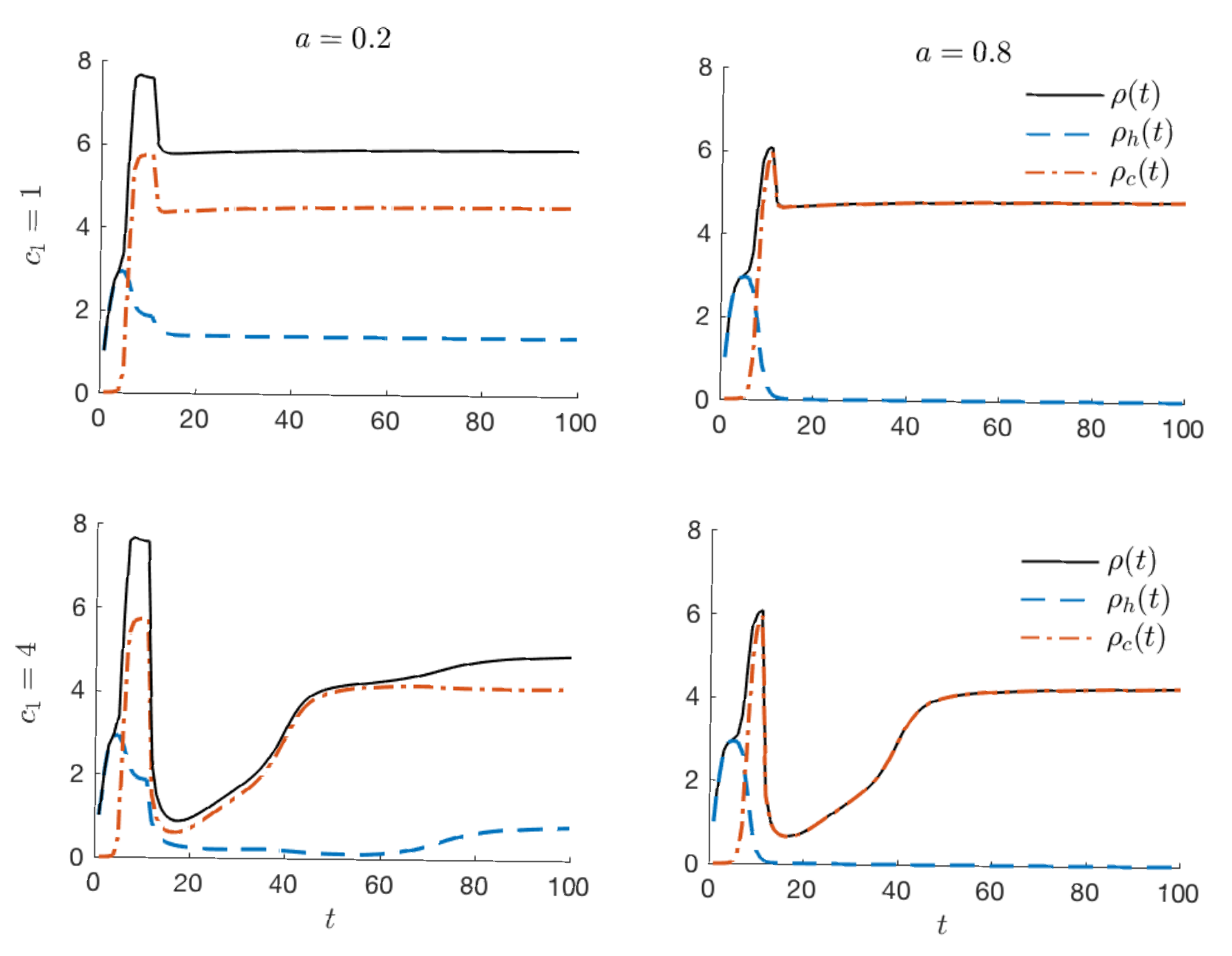} 
    }
    \centerline{     \rotatebox{90}{\hspace{2.4cm} \footnotesize Targeted }   
    \includegraphics[width=7.5cm]{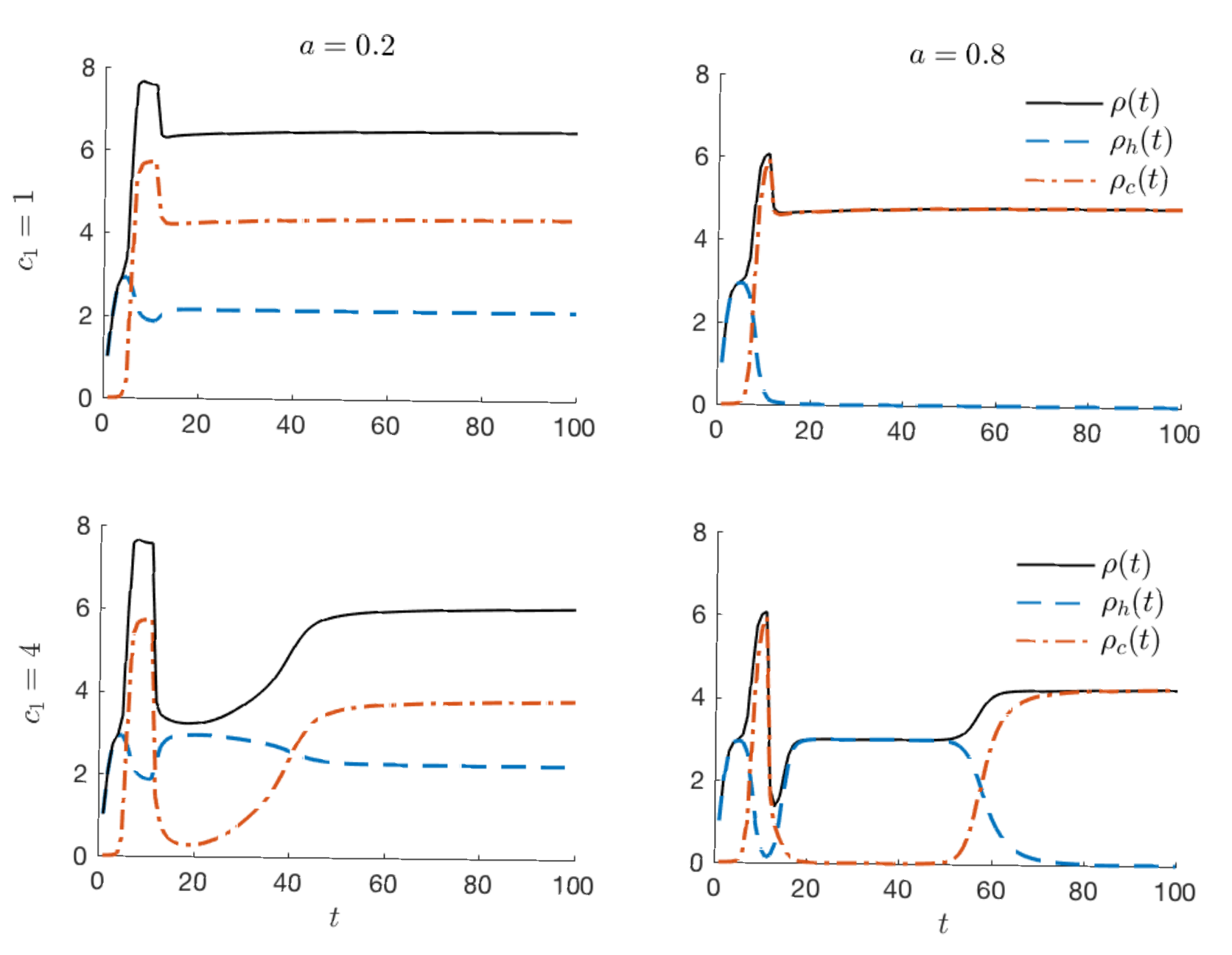} 
    }    
   \caption{Comparison of the total number of cells $\rho(t)= \rho_h(t)+\rho_c(t)$ when 
   using cytotoxic and targeted therapy with different values of the
   competition parameter $a = 0.2$ and $0.8$. 
   The drug is administered at $t_c=10$ with dosages $c_i = 1$ and $4$. 
   When $c_1=4$, administering a cytotoxic drug results with a relapse 
   regardless of the values of $a$. 
   The timing of the relapse is delayed in the case $a = 0.8$ when
   administering a high dosage $c_2=4$ of the targeted drug.
   }
\label{fig:4} 
\end{figure} 

We first examine the outcome of a single drug therapy 
for different values of the competition parameter $a$, using either a
cytotoxic or a targeted drug. 
We simulate the model 
without the diffusion term ($\nu_h = \nu_c = 0$).
For the initial condition, 
we set $\mu_1 = \mu_2 = 0$, $w = 10^{-5}$, and $\epsilon = 0.05$. 
The results shown in Figure~\ref{fig:4} confirm that the competition
parameter $a$ determines 
the outcome: either coexistence or aggressive competition.
This is the case with 
a low dosage $c_i=1$ as well as with a higher dosage $c_i=4$. 
When $a=0.2$, healthy cells and cancer cells are both present 
throughout the treatment, but not when $a=0.8$. 
In particular, under a high dosage of the cytotoxic drug, $c_1=4$,
the relapsed cancer cells overtake the population. 
In the case of a targeted drug, 
the healthy cells suppress the cancer cells for some time, so that the
relapse is delayed.  This does not prevent an eventual relapse.
The results are consistent with the observations of~\cite{Suijkerbuijk2016}
that showed that when the APC mutant clones in Drosophila midgut 
reach a certain size, they induce the apoptotic death of the surrounding wild-type cells. 
From this simulation, we observe that 
the targeted drug is partially effective in the competitive scenario, $a=0.8$.

\begin{figure}[!htb]
    \centerline{     
    \includegraphics[width=9cm]{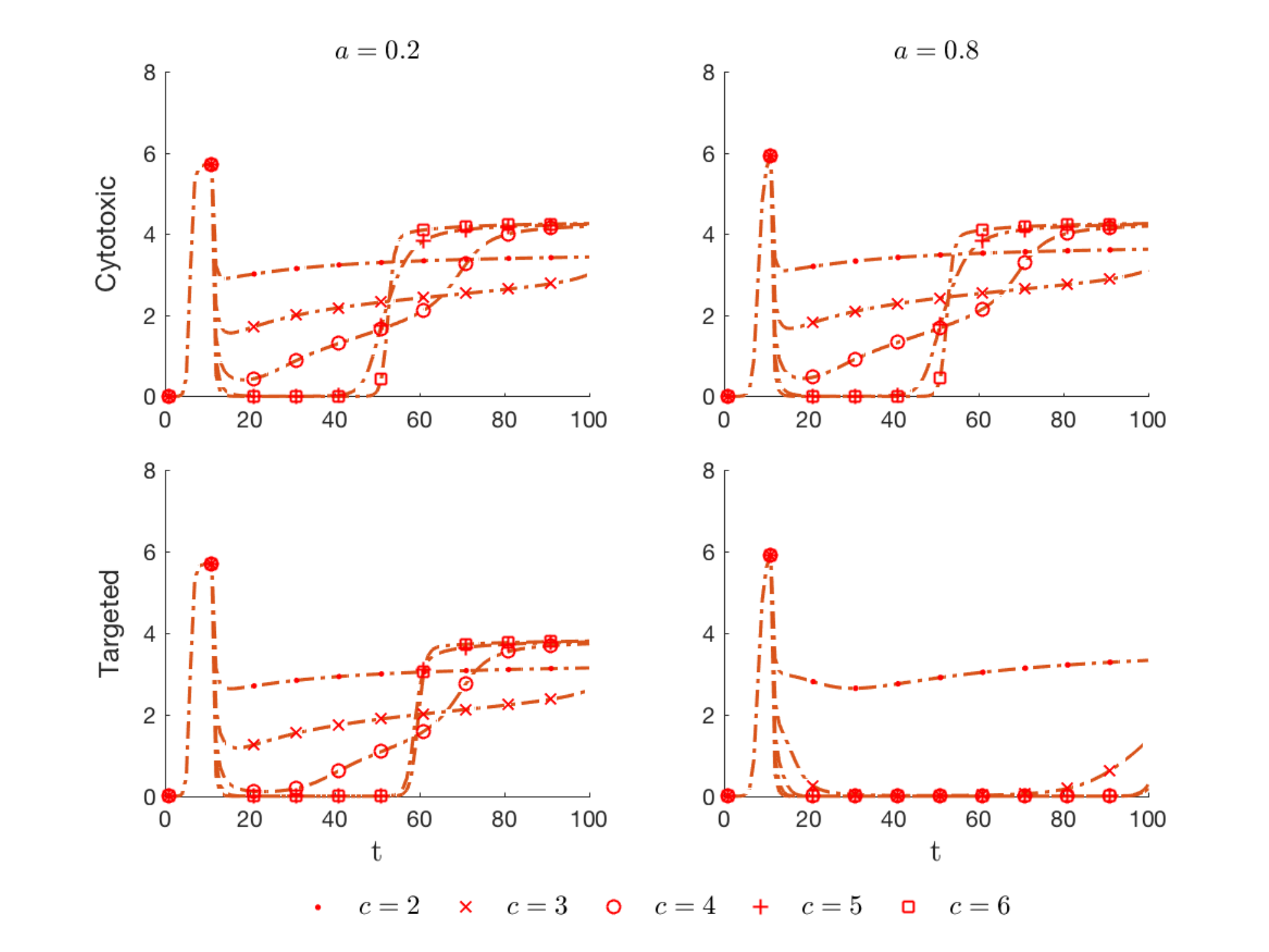} 
    }    
   \caption{Comparison of the total number of cancer cells $\rho_c(t)$ 
     for an increased drug dosage $2 \leq c \leq 6$. The results 
     compare cytotoxic therapy (up) and targeted therapy (down) for the
     competition parameters $a = 0.2$ (left) and $a=0.8$ (right). 
     We observe that in general, high cytotoxic drug dosages
     ($c=5,\,6$) result with a delayed, yet stronger relapse compared with 
     moderate dosages ($c=2,\,3$).   
     In contrast, targeted therapy results with a substantial delay in
     the relapse time in the highly competitive case $a=0.8$.
     }
\label{fig:5} 
\end{figure} 

\begin{figure}[!htb]
    \centerline{     
   \includegraphics[width=8cm]{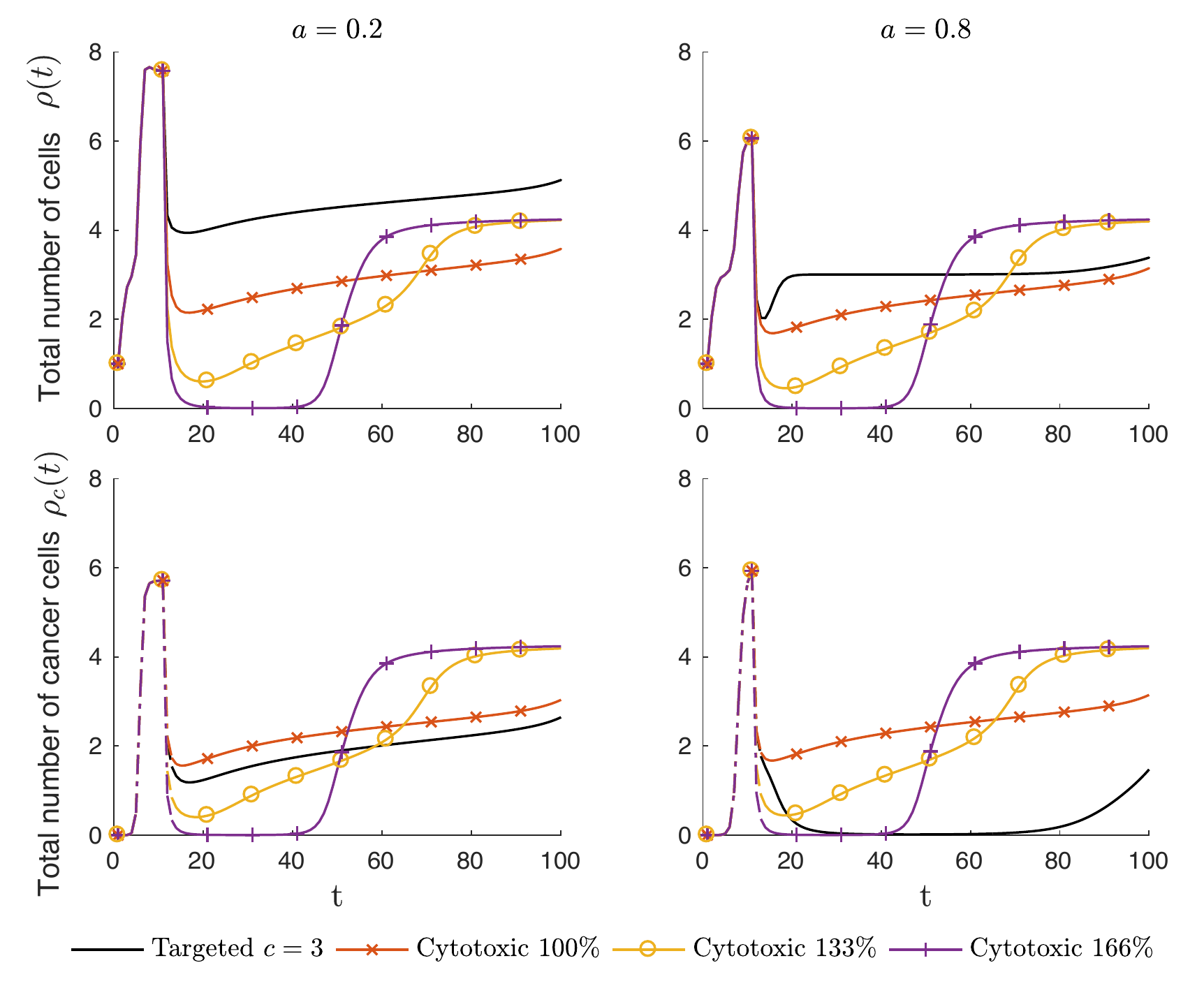} 
    }    
    \caption{Comparison of the total number of cells, $\rho(t)$ (top), 
      and cancer cells, $\rho_c(t)$ (bottom),
      using cytotoxic and targeted therapy. The uptake function of the
      cytotoxic drug is scaled such that it ranges from 100\% to 166\%
      of the uptake function of the targeted drug.
      The cytotoxic drug with a higher uptake reduces the cancer cells more effectively 
      than the targeted drug until a certain time, especially 
      when $a=0.2$. However, in the more competitive case, $a=0.8$,
      the targeted drug quickly becomes more effective. 
     }
\label{fig:6} 
\end{figure}

Figures~\ref{fig:5}--\ref{fig:6} show the effect of 
increasing the dosages in the range $2 \leq c \leq 6$ 
for the single drug therapy case, comparing 
cytotoxic and targeted drugs. 
Here, we assume less preexisting resistance by setting $\epsilon = 0.02$. 
In Figure~\ref{fig:5}, 
acute relapse is observed under cytotoxic drug 
with high dosages $c_i=5,\,6$,  
where the number of cancer cells rapidly increases 
at later times ($t>50$), compared with 
moderate dosages ($c=2,\,3$). 
This is the case for both competitive scenarios $a = 0.2$ and $0.8$. 
However, under targeted drugs, 
the relapse is worse when $a=0.2$, 
but not when the cells are highly competitive ($a=0.8$). 
The targeted drugs eliminate only the cancer cells which helps
the healthy tissue maintain its dominance and suppress 
the growth of cancer. In contrast, the cytotoxic drug, provides an advantage 
to the highly proliferative cancer cells, allowing them to fill in the void.

\begin{figure}[!htb]
    \centerline{  \rotatebox{90}{\hspace{0.6cm}\footnotesize   }    
   \includegraphics[width=9cm]{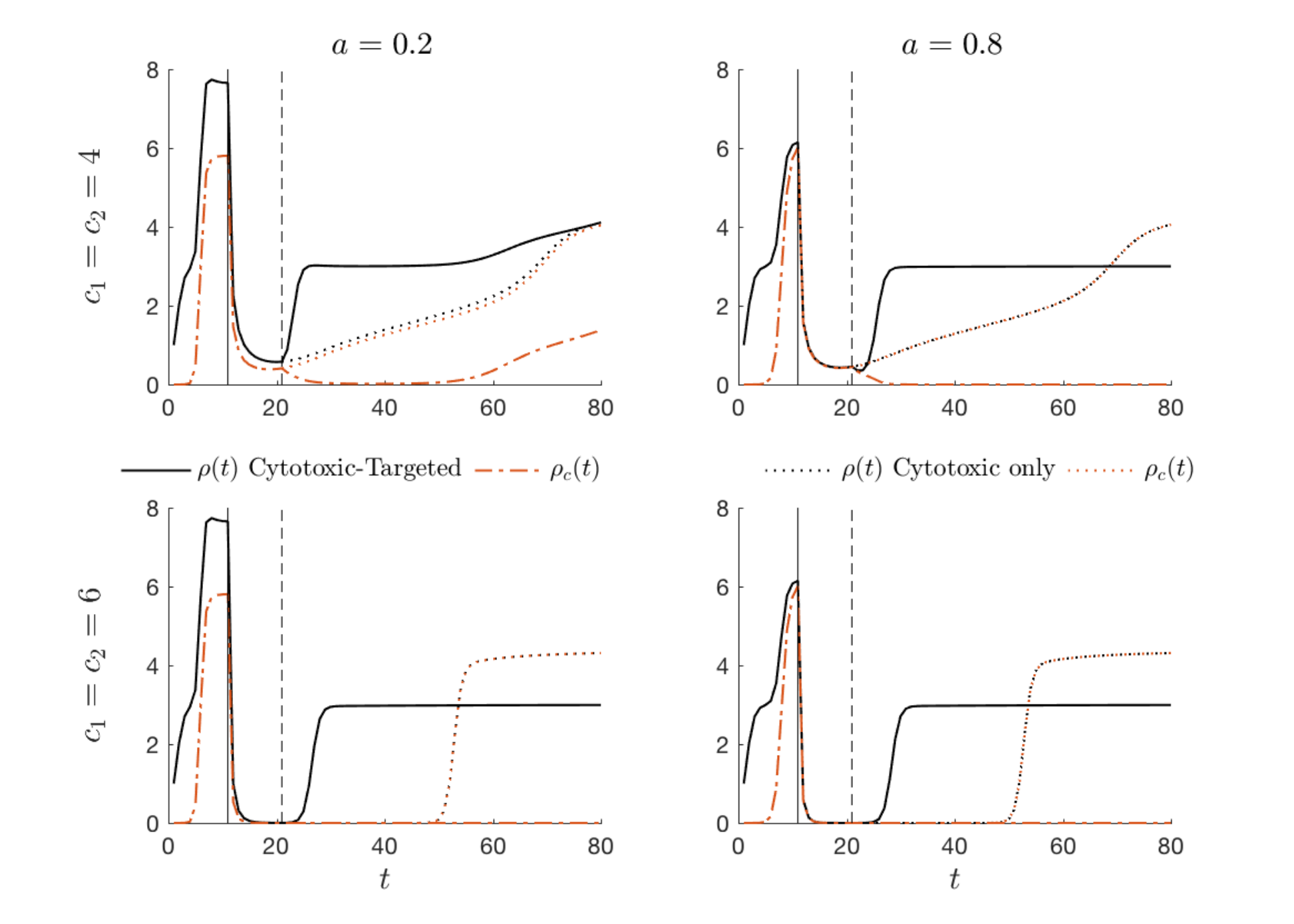} 
   }
    \centerline{  \rotatebox{90}{\hspace{0.6cm}\footnotesize   }    
   \includegraphics[width=9cm]{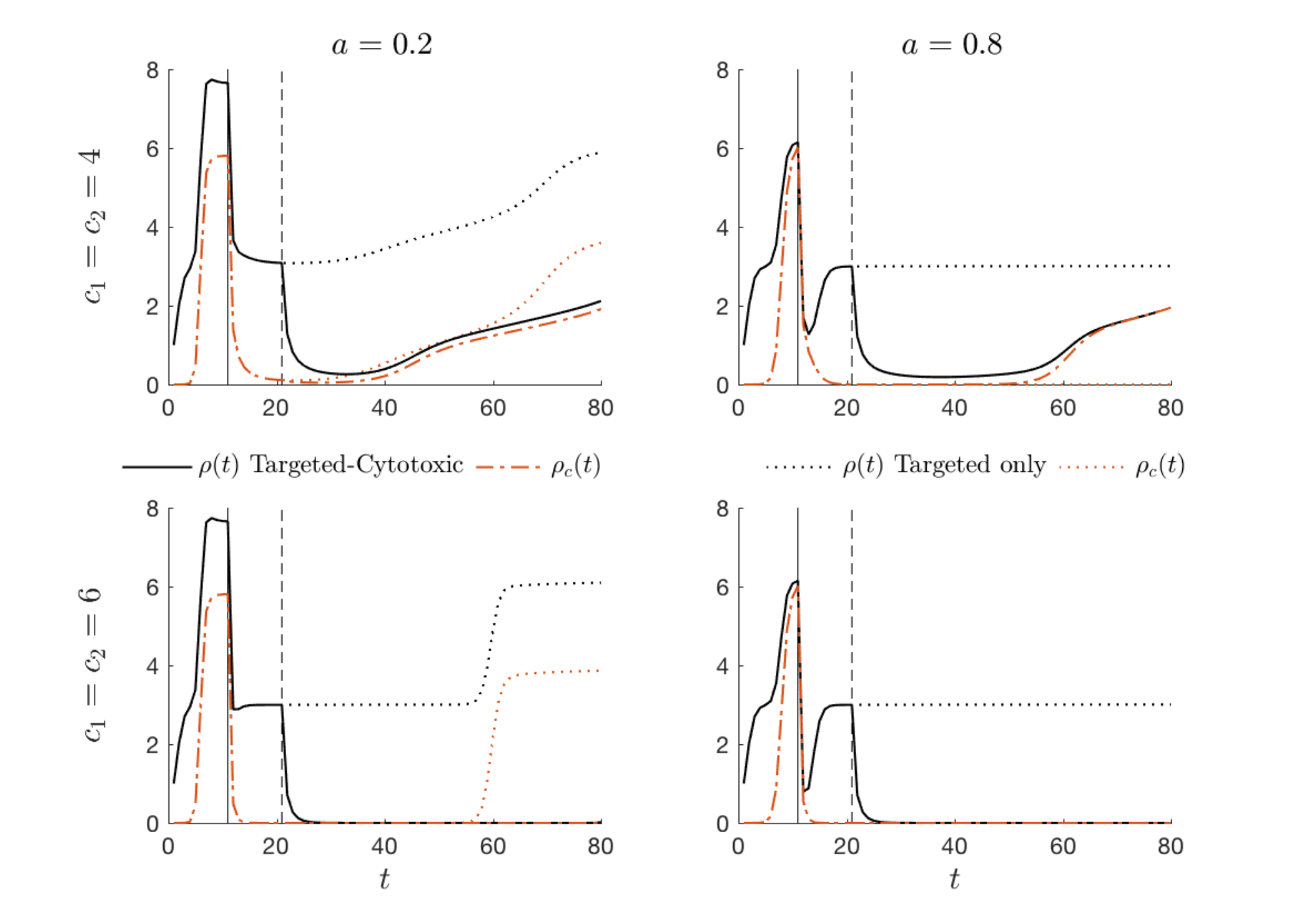} 
    }    
   \caption{The total number of cells $\rho(t)$ and cancer cells $\rho_c(t)$ 
   using a combination therapy, switching the drug once. 
   The initial drug is applied at $t=10$ and 
   switched to the second drug at $t=20$. 
   In the top 2 rows the cytotoxic drug is applied before the targeted
   drug.
   The order is reversed in the bottom 2 rows.
   Switching the drug in either way delays the relapse when $a=0.2$. 
   However, in the more competitive case, $a=0.8$, switching the targeted drug to a
   cytotoxic drug with insufficient amount of drug ($c_1=4$) results
   with a worse outcome for the cancer cells.
     }
\label{fig:7} 
\end{figure} 

We compare the results 
when the rate of apoptosis due to the cytotoxic drug 
is larger than the rate induced by the targeted drug, hoping that this
will provides insights about improved drug scheduling. 
Figure~\ref{fig:6} shows the results where 
the uptake function of the cytotoxic drug 
amplifies by up to 1.66 times the uptake function of the targeted drugs. 
The cytotoxic drug with a stronger uptake function is more efficient
in killing the cancer cells,
until a certain time point where 
the resistant cells cause a relapse. 
However, when $a=0.8$, 
stronger apoptosis of the cytotoxic drug holds 
for a very short period of time, so that 
the targeted drug has a significant advantage over the cytotoxic drug.
This suggests that there may exist a drug scheduling 
that maximizes the drug effect, particularly
when the competition is less aggressive ($a=0.2$).

To demonstrate the effectiveness of combination therapies,  
we compare the number of cells $\rho(t)$ and $\rho_c(t)$ 
using a single drug therapy, to 
a combination therapy switching either from a cytotoxic drug to a
targeted drug or the other way around.
We comment that often 
the first-line therapy is replaced by other drugs 
once it becomes ineffective 
\citep{Biswas2016, Mok2009, Kalemkerian2012}. 
Figure~\ref{fig:7} presents the results where 
the first drug is applied at $t=10$, and 
switched to the second drug at $t=20$. 
We set the values of the cytotoxic drug dosage, $c_1$, and the
targeted drug dosage, $c_2$, to either $4$ or $6$. 
We observe that switching the drug delays the relapse 
until the final simulation time $t=80$, in particular with the
higher drug dosage, $c_i = 6$. 
While a single drug therapy 
eventually results in a relapse 
due to the resistant population, 
we verify that 
switching the drug helps in delaying the relapse. 
This conclusion depends on the level of competition.
With a moderate dosage $c_i=4$, switching from a cytotoxic drug to a
targeted drug is effective when $a=0.2$, 
but not when $a=0.8$, where targeted drugs are advantageous.

\begin{figure}[!htb]
    \centerline{ 
       \includegraphics[width=8.5cm]{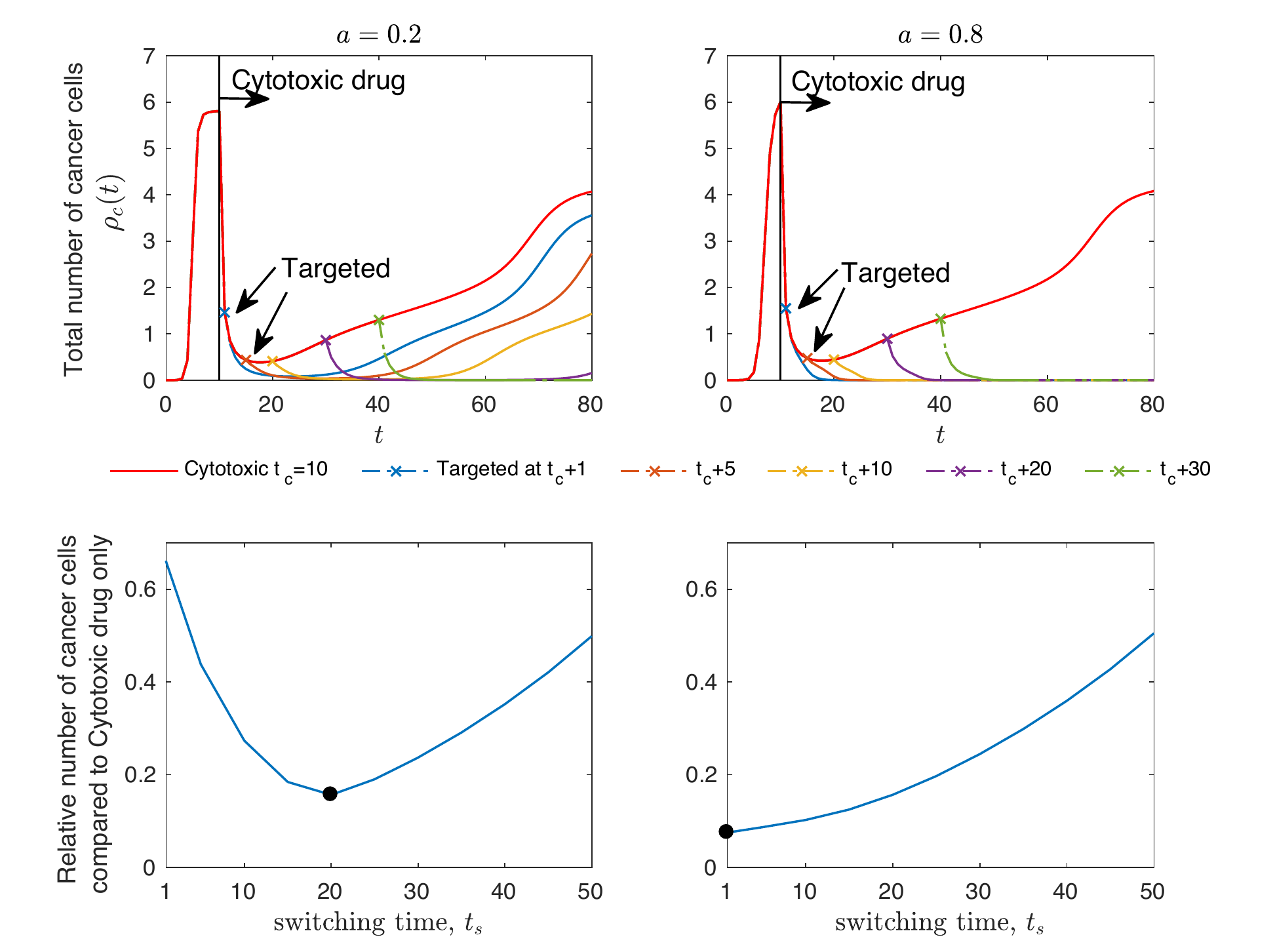} 
    }   
   \caption{Number of cancer cells when the treatment 
   begins with the cytotoxic drug at $t_c=10$, and
   then switches to the targeted drug at different times ($\times$). 
   The targeted drug delays the relapse due to resistance to the 
   cytotoxic drug.
   In particular, when $a=0.2$, there exists a switching time 
   that minimizes the number of cancer cells, approximately $t_c + 20$. 
   However, in the more competitive case, $a=0.8$, it is better to
   switch the drugs earlier 
   since the targeted drug is more effective. }
\label{fig:8-1} 
\end{figure}
\begin{figure}[!htb]
    \centerline{ 
       \includegraphics[width=8.5cm]{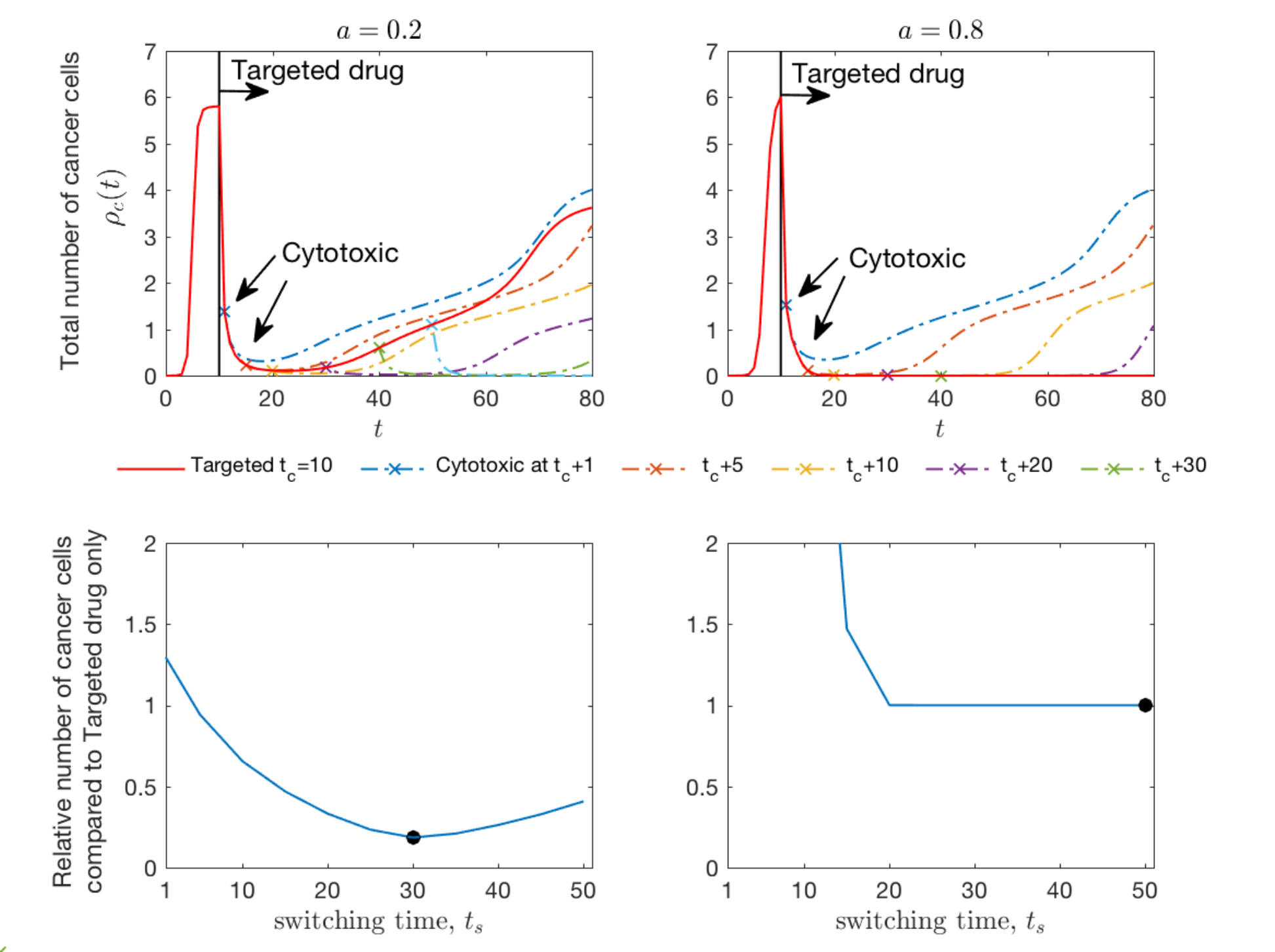} 
    }        
   \caption{Number of cancer cells when the treatment 
     begins with the targeted drug at $t_c=10$, and
     then switches to the cytotoxic drug at different times ($\times$). 
     The cytotoxic drug delays the relapse due to 
     targeted drug resistance when $a=0.2$. 
     Switching the drug at $t_c + 30$ 
     minimizes the total number of cancer cells. When $a=0.8$, 
     using the targeted drug only without switching to a cytotoxic
     drug is more effective. 
   } 
\label{fig:8-2} 
\end{figure}

We further aim to compute the optimal switching time. 
Figures~\ref{fig:8-1} and~\ref{fig:8-2} show
the total number of cancer cells, $\rho_c(t)$,
for the low-competition ($a=0.2$) and the highly competitive ($a=0.8$)
cases.  The drug is switched at different times after 
the initial drug is applied at $t_c = 10$. 
We fix the dosage as $c_i=4$. 
The case when the therapy is initiated with a cytotoxic drug 
is shown in Figure~\ref{fig:8-1}.
We observe that the cancer cell population remains low 
when the targeted drug is applied after the 
cytotoxic drug has eliminated sufficiently many
cells that were resistant to the targeted drug. 
The second row shows the relative size of the total 
cancer cell population compared to a single drug therapy, 
that is, $\left.\int_0^{80} \rho_c(t) \,dt \middle/ 
\int_0^{80} \rho^*_c(t) \,dt \right.$, where  
$\rho^*_c(t)$ is the number of cancer cells subject ot a 
cytotoxic drug treatment only. 
We observe that when $a=0.2$,  there exists an optimal switching time.
The relative cancer size is minimized  to 20\% 
when the therapy is switched at $t_c+20$.  
Figure~\ref{fig:8-2} shows the opposite case where 
the drug is switched from targeted to cytotoxic. 
Similarly, one can benefit from switching the drug 
when $a=0.2$. In this case, the optimal switching time is 
approximately $t_c + 30$. 
When $a=0.8$, a therapy using only the targeted drug with a dosage of $c_2 = 4$ 
is more effective than an alternating therapy.
As a result, rapidly switching the drug from targeted 
to cytotoxic yields significantly worse results, 
while for the switching from cytotoxic to targeted drug -- the 
sooner the better. 

\subsection{Continuous phenotypic levels of drug resistance}\label{sec:Num2} 

In this section, we study the implications of considering a 
continuous resistant space on the treatment scheduling.  Generally, we will observe 
variations in the mean resistant level as a function of the drug dosage. 
In a previous work we demonstrated that 
different continuum models of proliferation and uptake 
functions yield distinctive dynamics in the drug resistance space,
\cite{Cho2018MDR}.
The dynamics of continuum-resistance models can be similar to the dynamics
of models that are based on discrete levels of resistance.  It can
also be significantly different.
In linear models of proliferation $r(\theta)$ and 
drug uptake $\mu(\theta)$, the typical outcome is that cells end up
concentrating either in the most sensitive or in the most resistant trait.  Such
dynamics is essentially similar to considering a model with two
resistance states: fully resistant and fully sensitive.  Differences
between the continuum and discrete models are observed with non-linear
proliferation and drug uptake functions.
Here, we study how our model depends on the choice of continuum 
models 
by considering two functions:
(i) a quadratic model that allows 
intermediate resistance level,
$$r_c(\theta) = (\eta_c-\frac{\eta_c}{4}) {\prod_{i=1}^2 (1-\theta_i^2)} + \frac{\eta_c}{4},\,\, 
\mu_c(\theta) = \mu (\theta_1-1)^2, 
$$
\noindent and (ii) a linear model for which the outcome is similar to
a discrete two-states model,
$$r_c(\theta) = (\eta_c-\frac{\eta_c}{4}) {\prod_{i=1}^2 (1-\theta_i)} + \frac{\eta_c}{4},\,\, 
\mu_c(\theta) = \mu (\theta_1-1).
$$
The parameters are taken to be 
comparable with Eqs.~\eqref{eq:prolif}--\eqref{eq:uptake}. 
The death terms are assumed to be constant,
$d_h(\theta) = d_c(\theta) = 0.5$,
and the epimutation rates are set as $\nu_h = \nu_c = 10^{-3}$. 

\begin{figure}[!htb]
    \centerline{  \rotatebox{90}{\hspace{1.3cm} \footnotesize 
   Cytotoxic   }    
   \rotatebox{90}{\hspace{0.55cm} \footnotesize 
    $n_c(t,\theta) \hspace{1.0cm} n_h(t,\theta)$   }   
       \includegraphics[width=7cm]{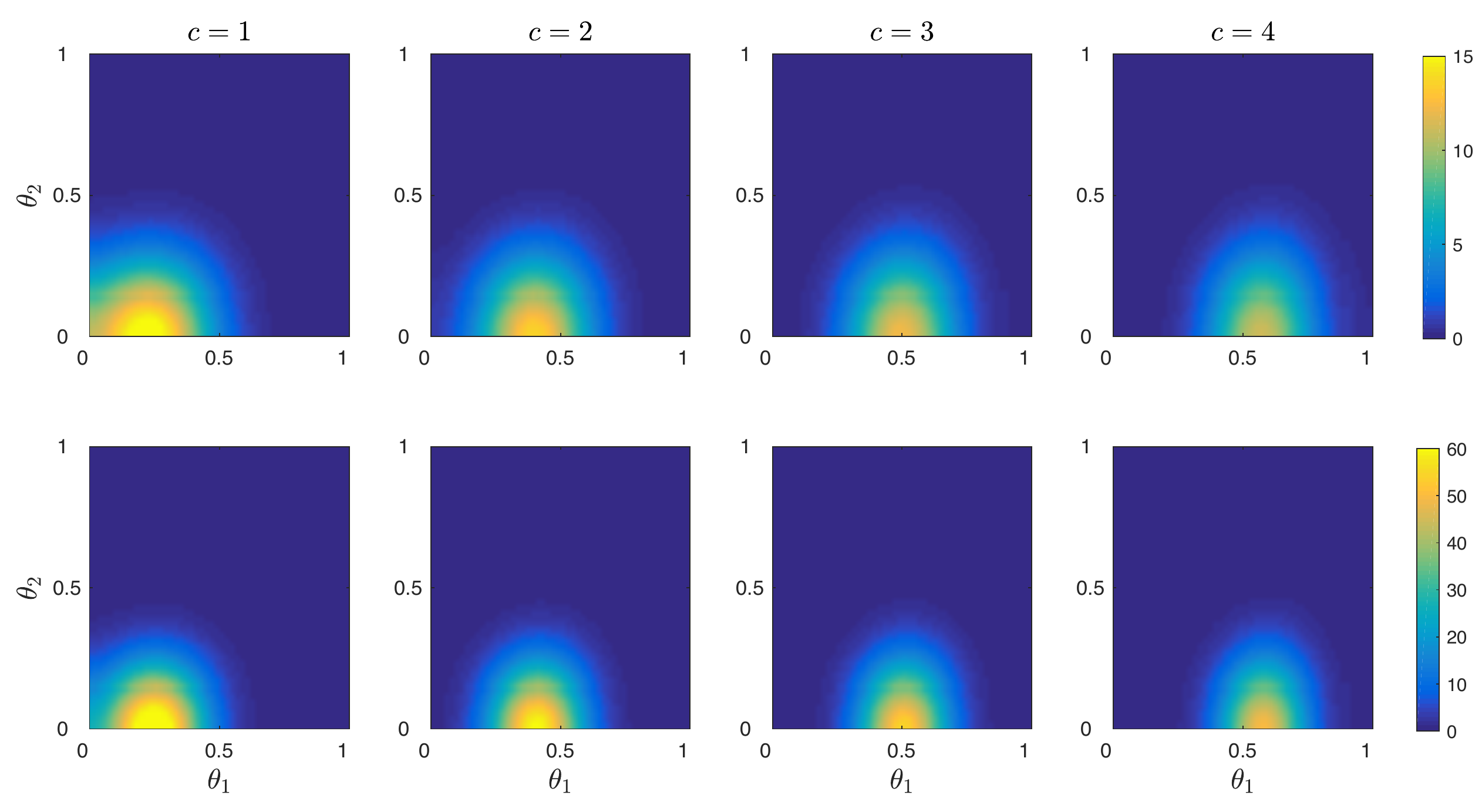} 
    }
    \centerline{  \rotatebox{90}{\hspace{1.3cm} \footnotesize 
   Targeted   }    
   \rotatebox{90}{\hspace{0.55cm} \footnotesize 
    $n_c(t,\theta) \hspace{1.0cm} n_h(t,\theta)$   }   
       \includegraphics[width=7cm]{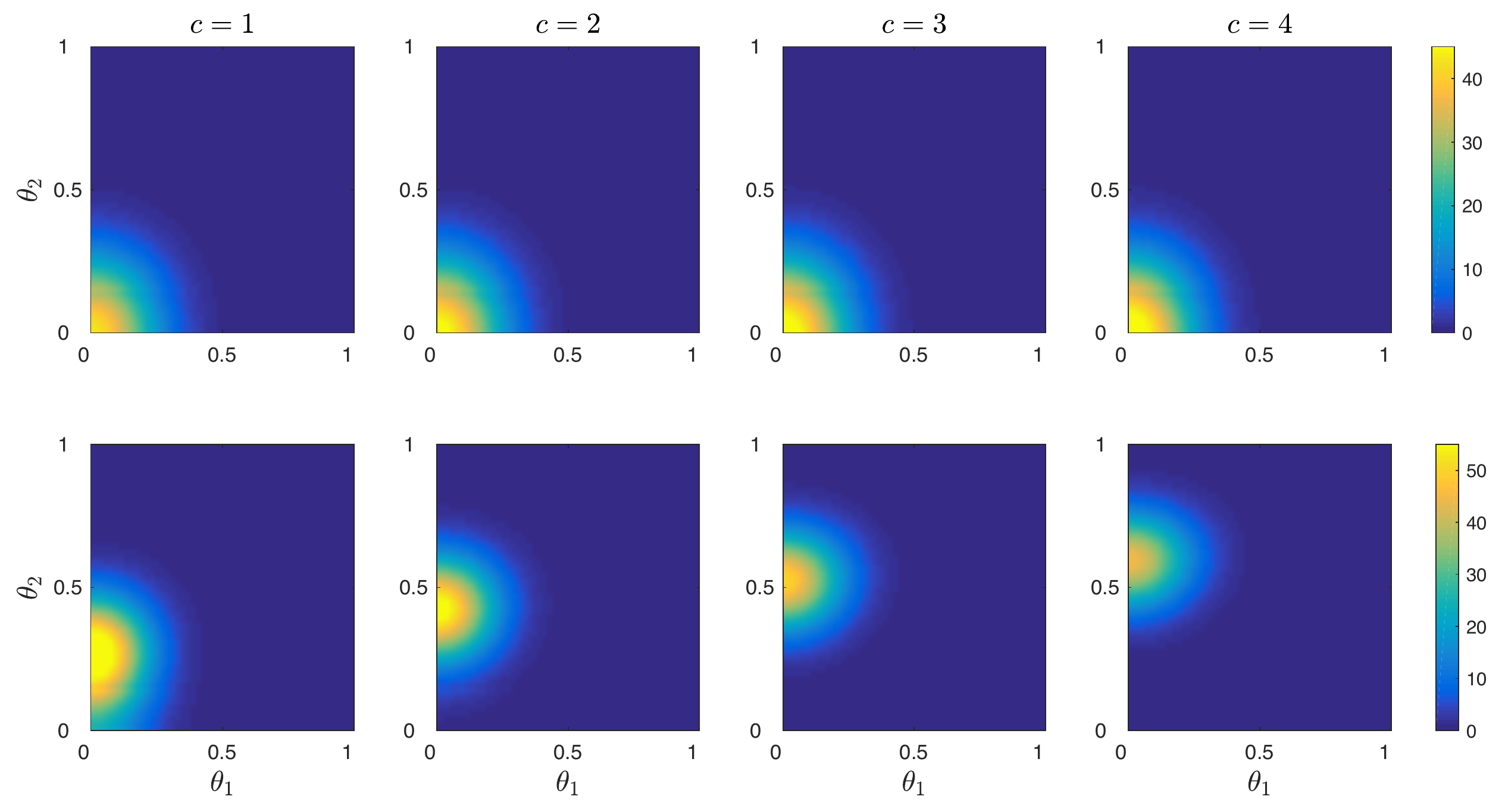} 
    }
   \caption{ Distribution of healthy cells, $n_h(t,\theta)$, and
     cancer cells, $n_c(t,\theta)$, 
     in the resistance phenotype space using a quadratic model  and $a=0.2$ 
   for different drug dosages $c = 1,\,2,\,3,\,4$ at $t=80$. 
   As the drug dosage increases, the mean resistance level in cancer cells gradually increases
   in the direction of $\theta_1$ or $\theta_2$ depending on the 
   drug type.   Healthy cells are only affected by the cytotoxic drug.
     }
\label{fig:11_1} 
\end{figure}

\begin{figure}[!htb]
    \centerline{  \rotatebox{90}{\hspace{1.3cm} \footnotesize 
   Cytotoxic   }    
   \rotatebox{90}{\hspace{0.5cm} \footnotesize 
    $n_c(t,\theta) \hspace{1.0cm} n_h(t,\theta)$   }   
       \includegraphics[width=7cm]{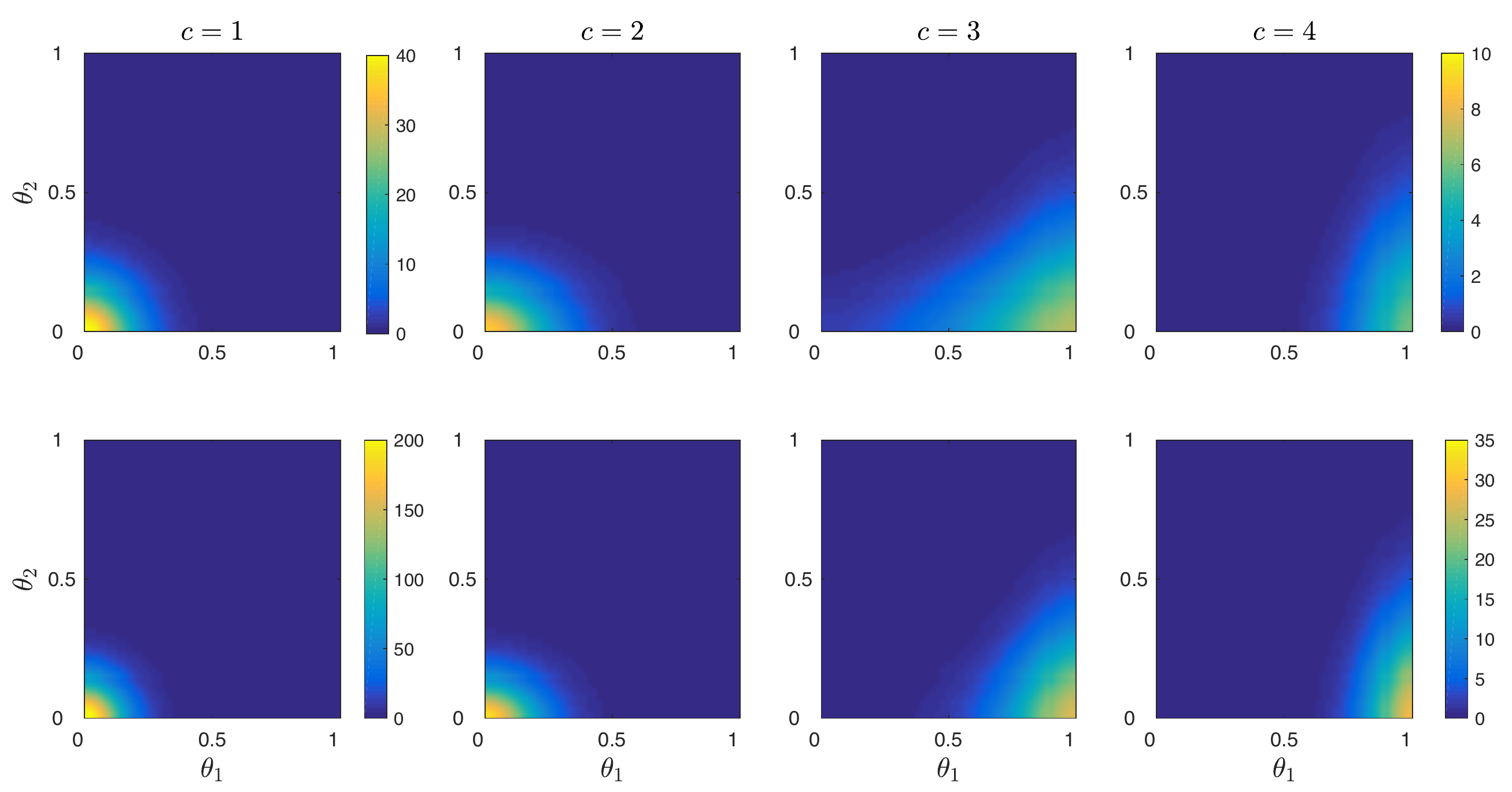} 
    }
    \centerline{  \rotatebox{90}{\hspace{1.3cm} \footnotesize 
   Targeted   }    
   \rotatebox{90}{\hspace{0.5cm} \footnotesize 
    $n_c(t,\theta) \hspace{1.0cm} n_h(t,\theta)$   }   
       \includegraphics[width=7cm]{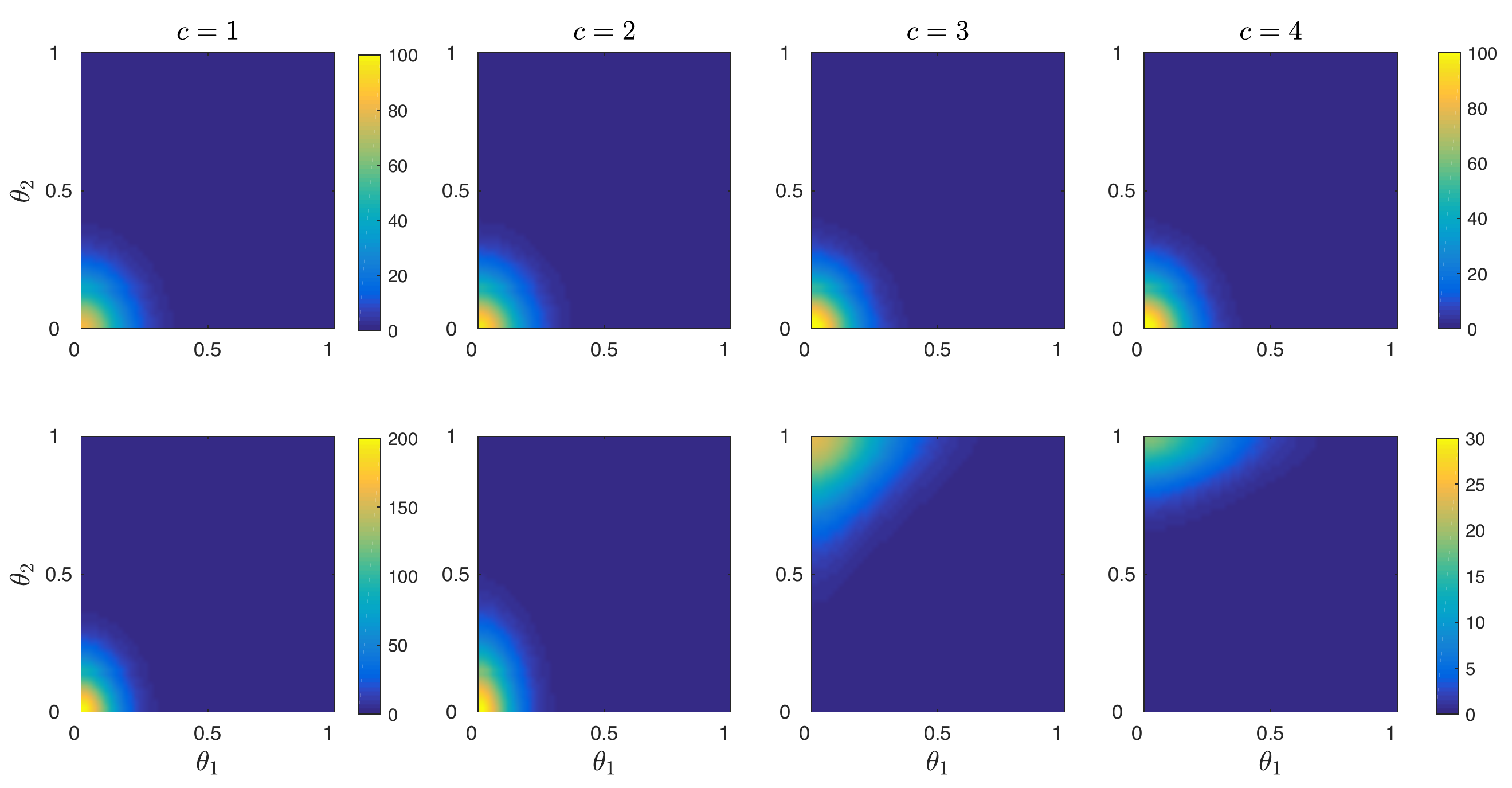} 
    }    
   \caption{ Distribution of healthy cells, $n_h(t,\theta)$, and
     cancer cells, $n_c(t,\theta)$, 
     in the resistance phenotype space using a linear model and $a=0.2$ 
     for different drug dosages $c = 1,\,2,\,3,\,4$ at $t=80$.  
     The level of resistance in cancer cells changes from fully-sensitive,
     $\theta_i = 0$, to fully-resistant, $\theta_i = 1$,
     depending on the drug dosage.  This is a sharp transition
     compared with the quadratic model in Figure~\ref{fig:11_1}. 
     }
\label{fig:11_2} 
\end{figure}

Figures~\ref{fig:11_1} and \ref{fig:11_2} show the distribution of both
healthy cells, $n_h(t,\theta)$,
and cancer cells, $n_c(t,\theta)$, in the resistance space.
These results are computed with the quadratic and linear models, 
while increasing the drug dosage, $c = 1,\,2,\,3,\,4$. 
The distributions are shown for the case when the competition allows 
cells to coexist ($a=0.2$ at time $t=80$). 
Figure~\ref{fig:11_1} 
confirms that the quadratic model allows for intermediate resistance levels 
that gradually increase with increased drug dosage. 
As the dosages of the cytotoxic and targeted drugs increase, 
the mean resistance level of cancer cells $n_c(t,\theta)$ 
increases in the corresponding direction of $\theta_1$ and $\theta_2$.
The resistance levels of healthy cells is affected only by
the cytotoxic drug.
In contrast to the smooth transition in the resistance level observed
in the quadratic model, the outcome of the linear model is 
closer to binary as shown in Figure~\ref{fig:11_2}. 
With the dosage threshold of $c \approx 3$,
the dominating resistance trait instantly changes 
from fully-sensitive ($\theta_i=0$) to fully-resistant ($\theta_i=1$). 
We note that in the highly competitive case, $a=0.8$, 
the distributions are similar to the results shown in 
Figures~\ref{fig:11_1} and \ref{fig:11_2}, only that the concentration
of healthy cells is relatively low, $n_h(t,\theta) \approx 0$.  

\begin{figure}[!htb]
    \centerline{   \rotatebox{90}{\hspace{1.3cm} \footnotesize 
   model 1 }    
       \includegraphics[width=7cm]{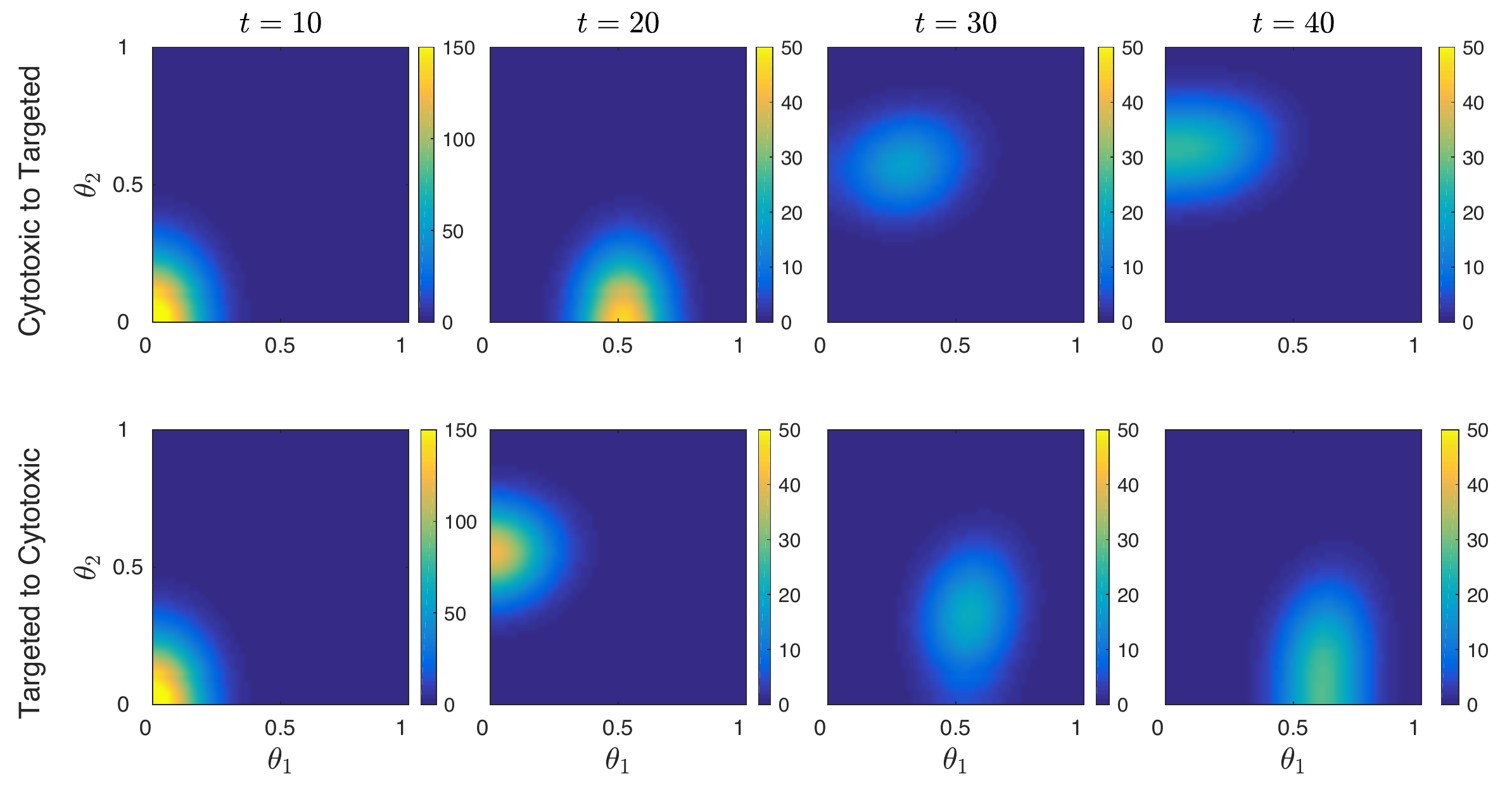} 
    }
    \centerline{   \rotatebox{90}{\hspace{1.3cm} \footnotesize 
   model 2 }    
       \includegraphics[width=7cm]{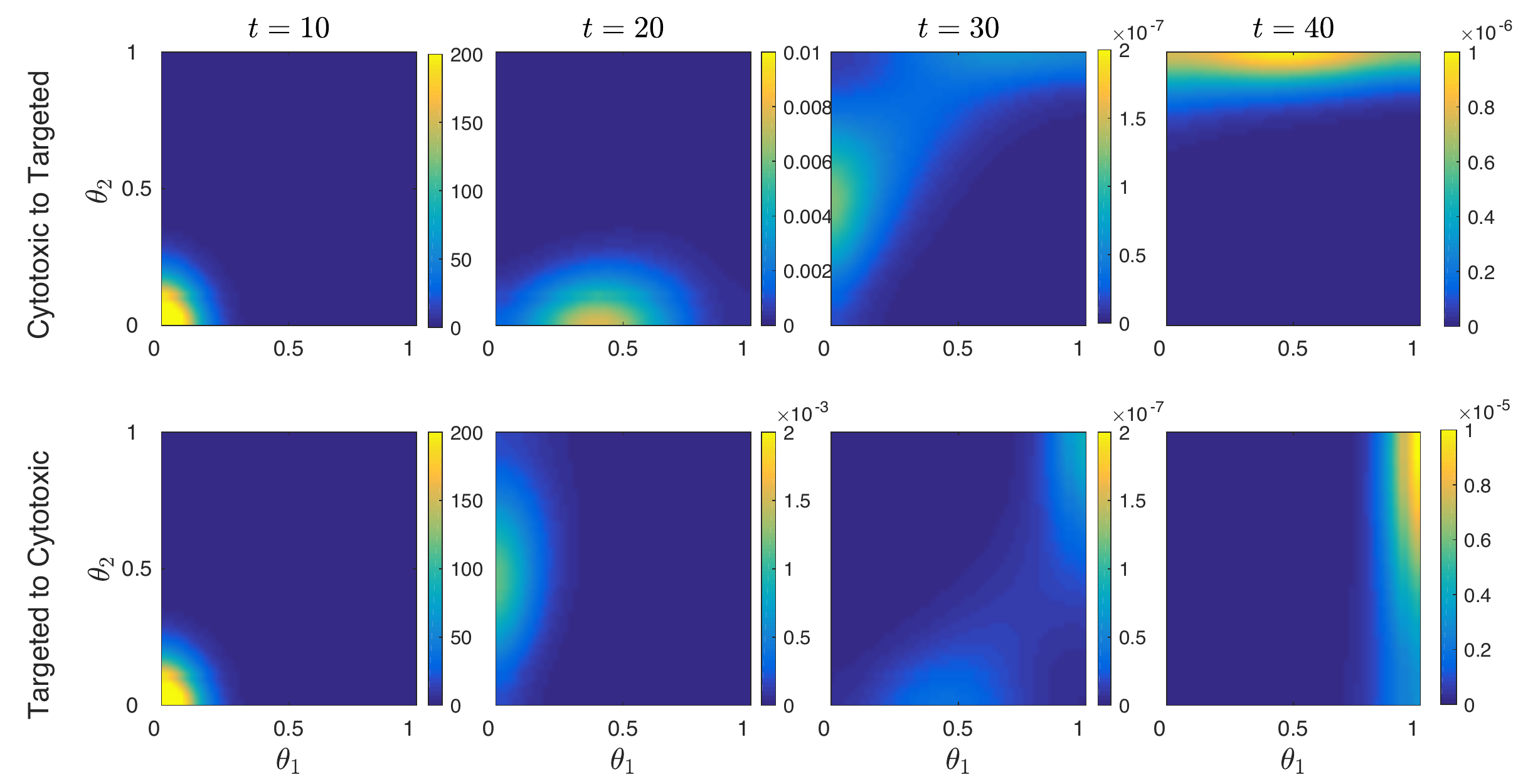} 
    }            
   \caption{ Evolution of the cancer cell distribution 
   $n_c(t,\theta)$ at $t=10,\,20,\,30,\,40$ 
   using the quadratic model (top) and the linear model (bottom) with
   a drug switching therapy. The dosages satisfy $c_1 + c_2 = 5$. 
   The initial drug is applied at $t=10$, and the second drug is applied
   at $t=20$.
   In contrast to the
   single drug therapy, a combination therapy reduces the levels of
   cancer cells that are resistant to one drug yet are 
   sensitive to the other drug.
 }
\label{fig:15_2} 
\end{figure}

We proceed to studying the evolution of the distribution in the
resistance space under drug switching therapy.  The results are shown
in Figure~\ref{fig:15_2}. 
Here, the initial drug is applied at 
$t_c = 10$ and is then switched to the second drug at $t=20$.
Prior to the treatment, 
the most sensitive cells dominate the population. 
However, after treatment is initiated, 
the resistant cells emerge, depending on the drug type. 
For instance, when the therapy is switched from a
cytotoxic drug to a targeted drug, 
the distribution shifts from $\theta_1\gg 0$ and $\theta_2=0$ 
to $\theta_2\gg 0$. 
Compared with the outcome of a single drug therapy, 
the population of cancer cells that are resistant to one drug 
and sensitive to the other drug declines. 
Cells that are resistant to both drugs, $\theta_1\approx 1$ and
$\theta_2\approx 1$, are likely to survive.  

\begin{figure}[!htb]
    \centerline{  \rotatebox{90}{\hspace{0.6cm}  }    
       \includegraphics[width=7cm]{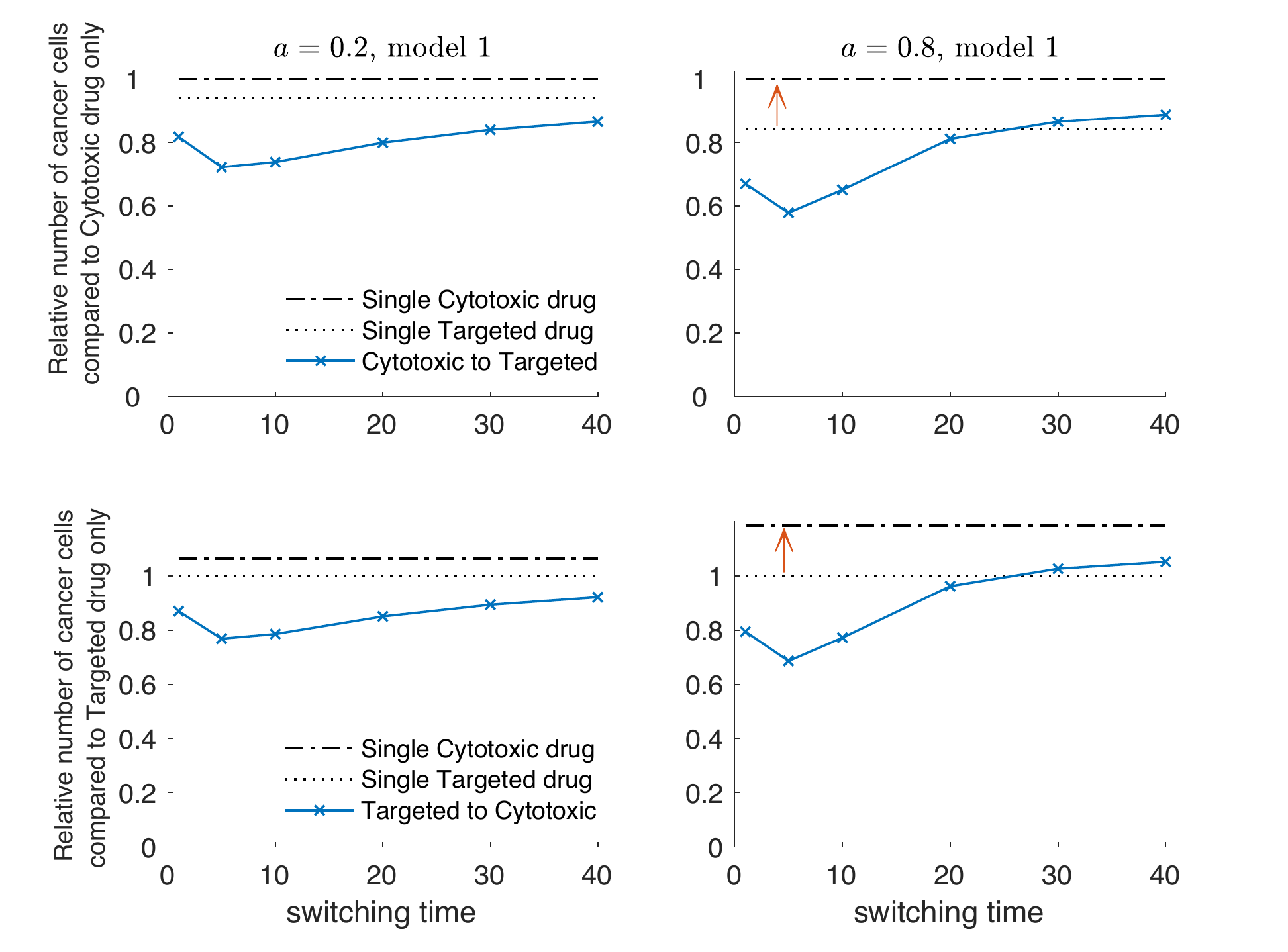} 
    }
    \centerline{  \rotatebox{90}{\hspace{0.6cm}  }    
       \includegraphics[width=7cm]{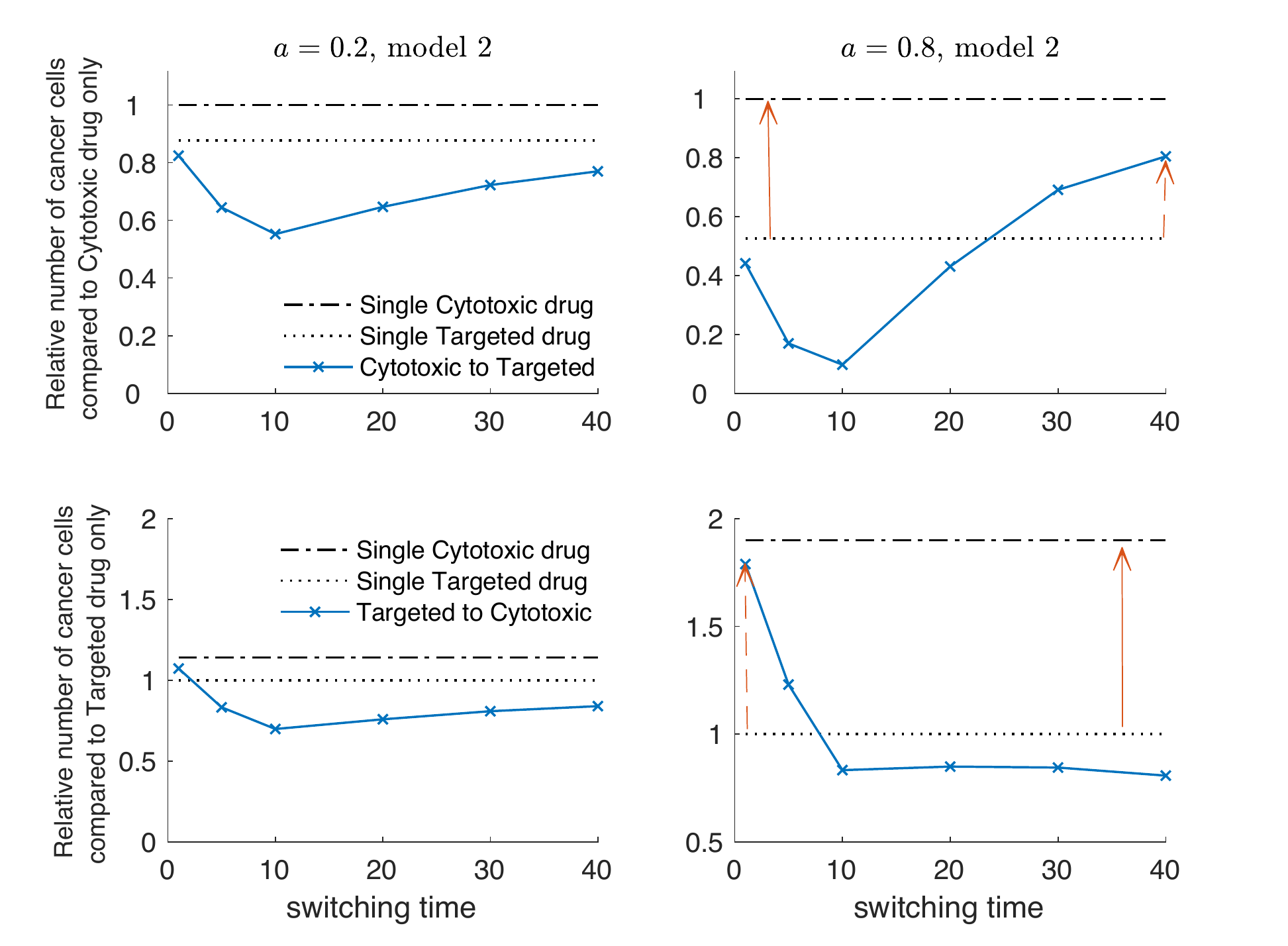} 
    }
   \caption{ 
   Relative number of cancer cells using drug switching therapy 
   compared to using a single cytotoxic drug  
   for different values of drug switching time $t_s$. 
   The results are shown for the quadratic model (top) 
   and the linear model (bottom). 
   The linear model is more sensitive to the switching time, 
   and when $a=0.8$, the outcome of 
   a single cytotoxic drug therapy or the
   switching therapy, can be
   worse than a single targeted drug therapy, 
   where the arrows indicate the increased amount. 
     }
\label{fig:13} 
\end{figure}

Figure~\ref{fig:13} compares the effect of switching therapy 
to a single drug therapy for the two continuum models. 
The relative size of the total 
cancer cell population compared with a single (cytotoxic) drug therapy, 
is $\left.\int_{10}^{100} \rho_c(t) \,dt \middle/ 
\int_{10}^{100} \rho^*_c(t) \,dt \right.$, where  
$\rho^*_c(t)$ is the number of cancer cells
under a cytotoxic drug treatment.
Once again, the results confirm that a single targeted drug therapy is 
particularly effective 
when the cells are highly competitive, $a = 0.8$. 
This is more significant in the linear model, 
where the relative size of the cancer cell population reduces
approximately to 50\%, compared with 80\% with the quadratic model.
When $a=0.2$, 
the switching therapy in any order 
is more effective than the single drug therapies regardless 
of the continuum model ($t_s \geq 5$). 
We note that the outcome of linear model
strongly depends on the switching time $t_s$, 
and particularly becomes worse 
than the single targeted drug therapy 
when $a=0.8$, if the tumor is not exposed 
to the targeted drug for a sufficiently long period. 

\subsection{Alternating therapies and on-off combination therapies}\label{sec:Num3}

\begin{figure}[!htb]
    \centerline{  \rotatebox{90}{\hspace{2.9cm} \footnotesize $c_1+c_2 = 3$  }  \hspace{.2cm}       
       \includegraphics[width=6.9cm]{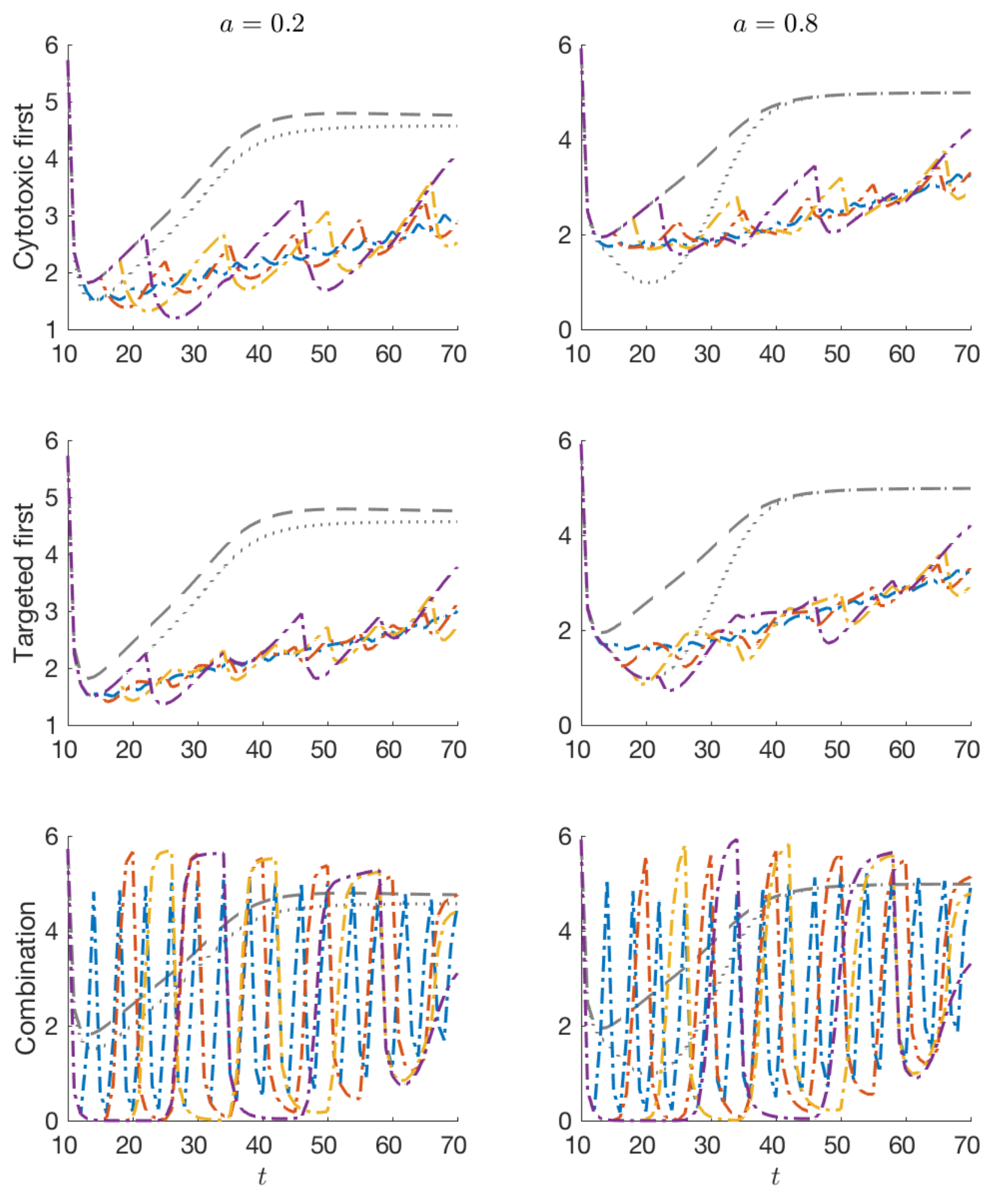} 
    } 
    \centerline{  \hspace{0.4cm}
       \includegraphics[width=6.5cm]{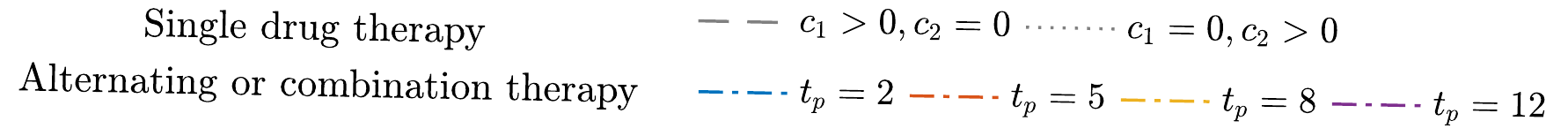} 
    }
   \caption{ The number of cancer cells $\rho_c(t)$ 
   using different therapies including single-drug, alternating, and combination therapy with a relatively low dosage $c_1 +c_2 = 3$. 
   The periods of alternating and combination therapies are taken 
   as $t_p=2$, $5$, $8$, and $12$. 
   The alternating therapy in any order is more effective than others 
   considering the overall number of cancer cells during the treatment. 
   A shorter alternating period ($t_p=2$) suppresses the cancer cells
   without oscillations.
   On the other hand, the on-off combination therapy yields a highly 
   oscillatory outcome. 
 }
\label{fig:16} 
\end{figure}

\begin{figure}[!htb]
    \centerline{ \rotatebox{90}{\hspace{2.9cm} \footnotesize $c_1+c_2 = 5$  }  \hspace{.2cm}     
       \includegraphics[width=6.8cm]{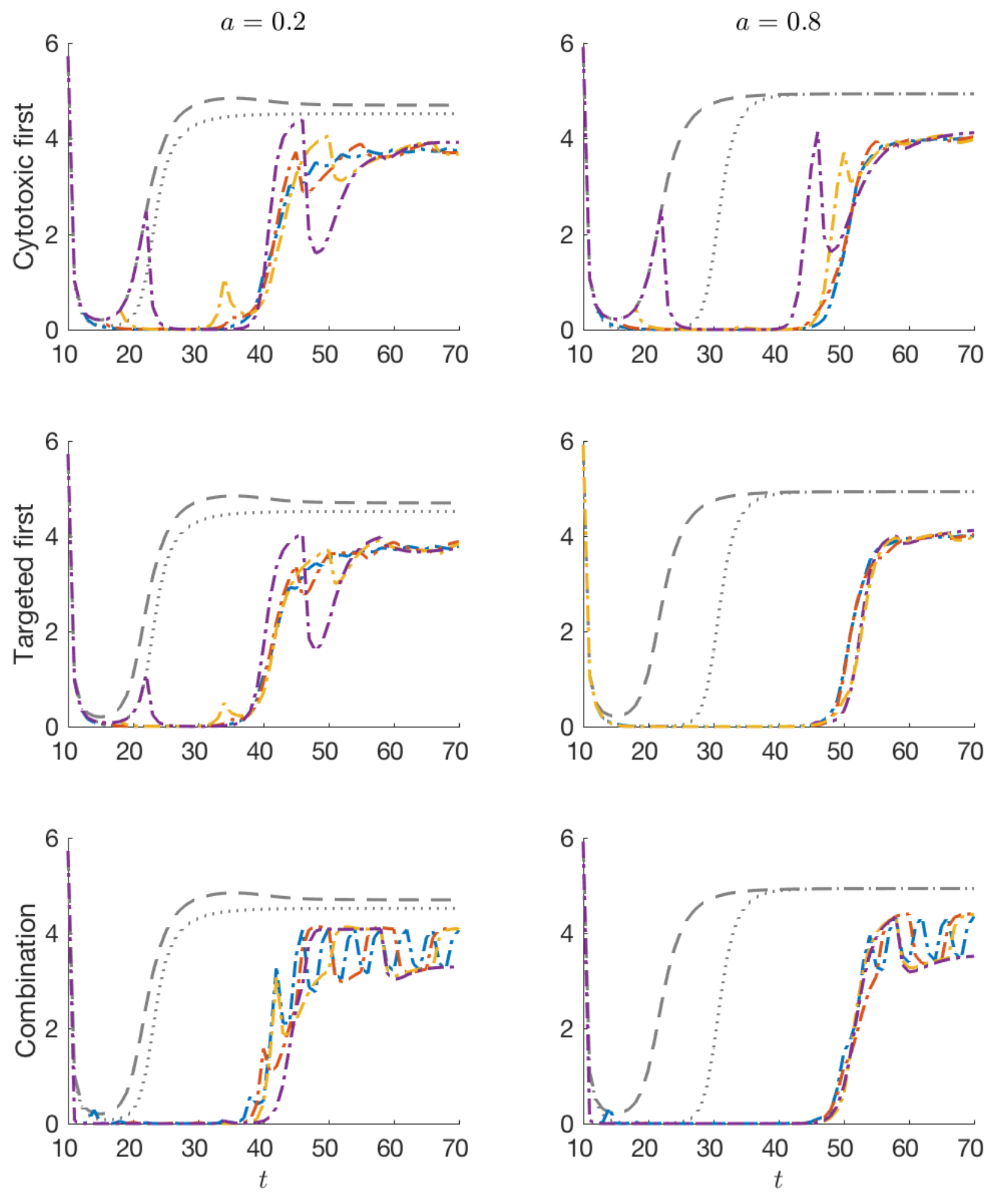} 
    }   
    \centerline{  \hspace{0.4cm}
       \includegraphics[width=6.5cm]{180514_nTotC_SwitchallP_ncase4_legend_.pdf} 
    }
   \caption{The number of cancer cells $\rho_c(t)$ 
   using different therapies including single-drug, alternating, and 
   combination therapy with a relatively high dosage $c_1 +c_2 = 5$, 
   and period $2$, $5$, $8$, $12$. We observe that 
   alternating the drug as quickly as possible ($t_p=2$) 
   delays the relapse while 
   suppressing the cancer cells.  In case of $a=0.8$, 
   initiating the therapy with the targeted drug 
   is more effective than initiating it 
   with the cytotoxic drug. The on-off combination therapy also delays the relapse 
   effectively. }
\label{fig:17} 
\end{figure}

In the previous sections we studied the emerging dynamics when
switching cytotoxic and targeted drugs.  The study was performed
assuming competition between healthy cells and cancer cells, and a
continuum resistance trait.
In this section, we assume that the drug can be changed 
within a short period of time, and
study the periodically alternating therapy and the on-off combination
schedules depicted in Figure~\ref{fig:c(t)}(c,d). 
The results for these studies are shown in
Figures~\ref{fig:16}--\ref{fig:17}.
In both figures, the drug therapies start at $t_c = 10$.
We test for different dosages: (i) a moderate dosage $c_1+c_2=3$ 
aiming at maintaining the cancer cell population at low levels; and (ii) a high
dosage, $c_1+c_2=5$, aiming at completely eliminating the cancer
cells.  We also test for different competition rates $a=0.2$ and $0.8$. 
As a reference, we plot the results obtained with single drug therapy.

Figure~\ref{fig:16} shows the dynamics of the cancer cells
under the moderate dosage.
We observe that the alternating therapy is more effective than the
other therapies, as it reduces the overall number of cancer cells
compared to single drug therapies. 
In particular, alternating the drug with a short period ($t_p=2$) 
suppresses the cancer growth without oscillations.
Initiating the alternating therapy in any order ends with similar results 
for both competition rates
since the preexisting resistant populations
to both drugs are identical.
On the other hand,   
the on-off combination therapy yields a highly 
oscillatory behavior that results in larger numbers 
of cancer cells during the off period compared with
the single drug therapy.
Since the dosage of chemotherapy is restricted 
due to its toxic nature \citep{Dorris2017,Ribeiro2012}, 
the on-off combination therapy 
may not be effective in such situations.

In the case of a higher dosage, $c_1+c_2=5$, 
we observe that the number of cancer cells reduces to 
less than an order of magnitude for a certain time period 
before a relapse occurs. 
As shown in Figure~\ref{fig:17}, both of the alternating  
and combination schedules significantly delay the relapse compared 
to the single drug therapies. 
As before, alternating the drug with shorter periods ($t_p=2$) 
keeps the cancer population below a certain threshold, 
in contrast to longer periods that may result in 
small peaks throughout the treatment. 
However, when $a=0.8$, we observe that initiating 
the therapy with the targeted drug is more effective 
in the sense that it overcomes the drawback of 
longer drug periods. 
We also observe that the on-off combination therapy with a high dosage 
effectively reduces the cancer cells population and delays the relapse, 
similarly to the alternating schedule with short periods. 
However, after the relapse, the cancer cell population fluctuates more
than with the alternating schedule. 

The results obtained so far assume equal sizes of preexisting 
populations that are resistant to the cytotoxic and to the targeted drugs. 
However, since the preexisting resistance is one of the critical 
factor in relapse, we further study the more realistic scenario in
which different fractions of the pre-treatment population are
resistant to both drugs.
We denote the number of cells that are resistant to the
cytotoxic drug and to the targeted drug as 
$$\rho_{c,R1}(t) = \int_{\bm 1_{\theta_1 \geq 0.5}} \hspace{-.3cm} n_c(t,\theta) d \theta,\quad \rho_{c,R2}(t) = \int_{\bm 1_{\theta_2 \geq 0.5}} \hspace{-.3cm} n_c(t,\theta) d \theta,$$  
respectively. We remark that  
the previous results correspond to the case 
$\rho_{c,R1}(0) = \rho_{c,R2}(0)$. 
We consider the initial condition as in Eq.~\eqref{eq:IC}, 
but with different variances $\epsilon = 0.02$ or $0.08$ 
for each direction, $\theta_1$ or $\theta_2$. 
The ratio then either becomes
$\rho_{c,R1}(0) : \rho_{c,R2}(0) = 1 : 10^4$
or $\rho_{c,R1}(0) : \rho_{c,R2}(0) = 10^4 : 1$.

\begin{figure}[!htb]
    \centerline{  \rotatebox{90}{\hspace{2.9cm} \footnotesize $c_1+c_2 = 3$  }  \hspace{.2cm}    
       \includegraphics[width=7cm]{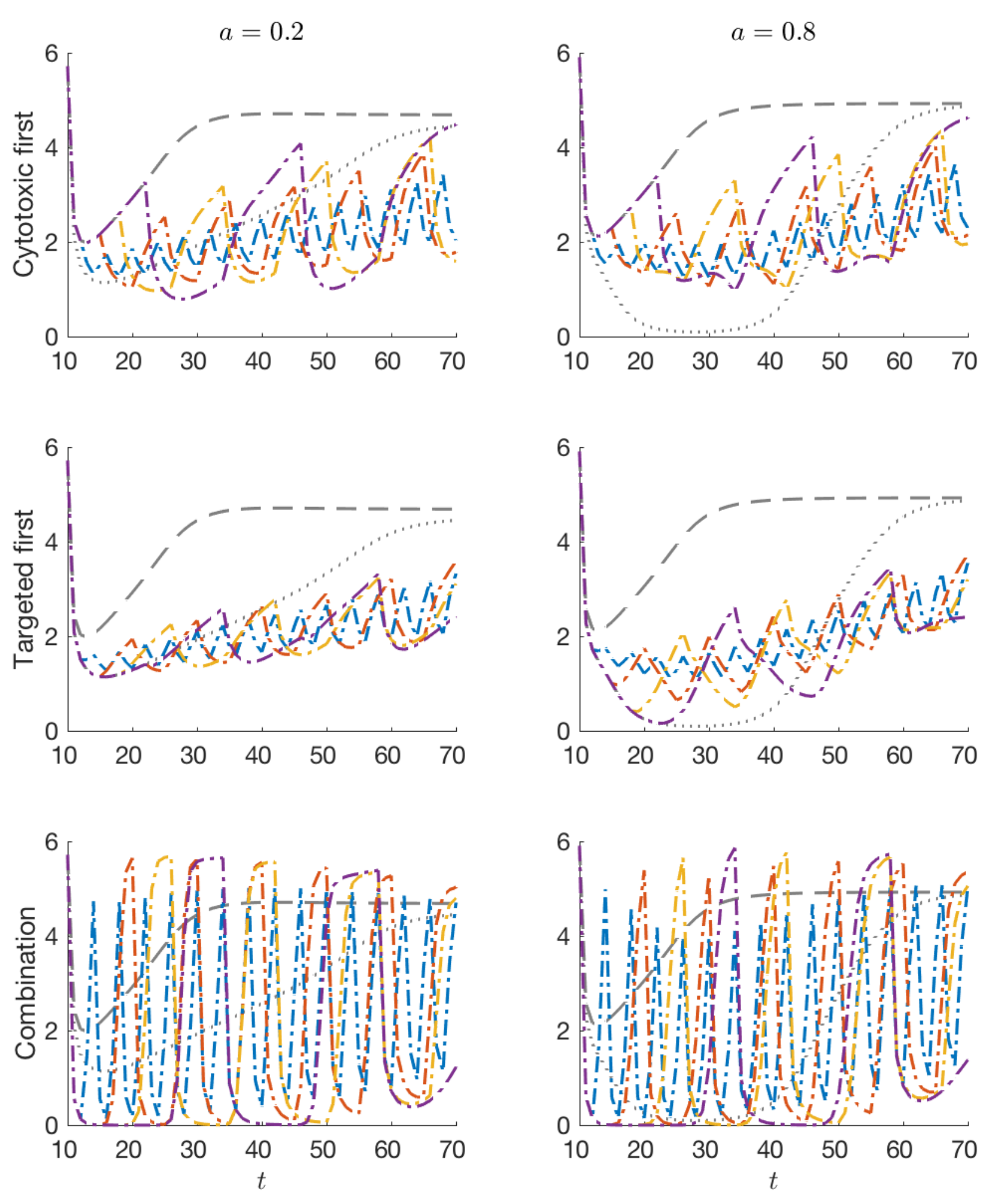} 
    }            \vspace{.1cm} 
    \centerline{  \rotatebox{90}{\hspace{2.9cm} \footnotesize $c_1+c_2 = 5$  }    \hspace{.2cm}    
       \includegraphics[width=7cm]{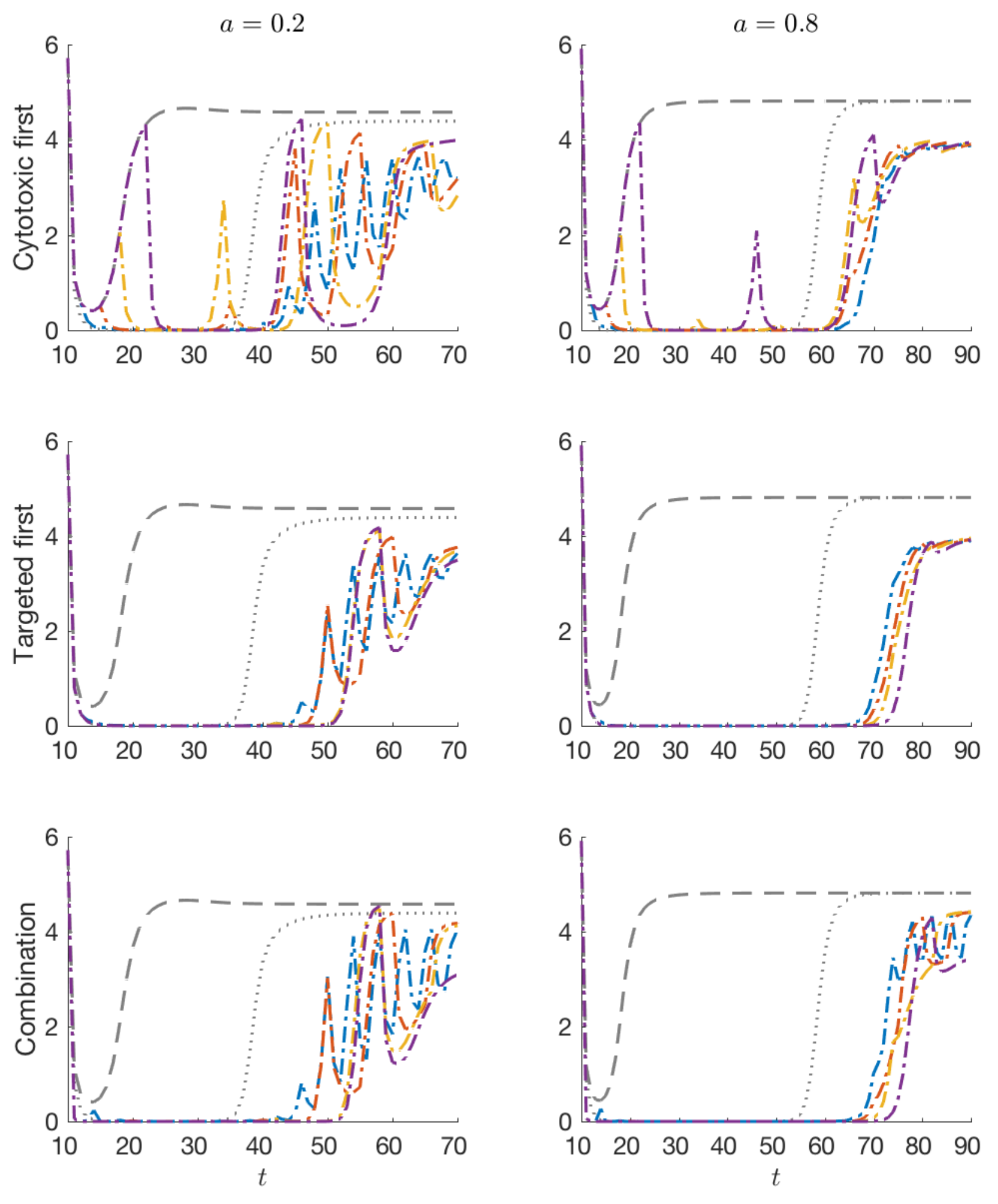} 
    }           
    \centerline{  \hspace{0.4cm}
       \includegraphics[width=6.5cm]{180514_nTotC_SwitchallP_ncase4_legend_.pdf} 
    }
   \caption{Number of cancer cells $\rho_c(t)$ for different therapies when 
   the number of cells that are resistant to the cytotoxic drug is larger than those 
   that are resistant to the targeted drug ($\rho_{c,R1} > \rho_{c,R2}$). We observe that initiating 
   the alternating therapy with the targeted drug with a smaller resistant population 
   is more effective.   
     }
\label{fig:18} 
\end{figure}

\begin{figure}[!htb]
    \centerline{  \rotatebox{90}{\hspace{2.9cm} \footnotesize $c_1+c_2 = 3$  }  \hspace{.2cm}      
       \includegraphics[width=7.15cm]{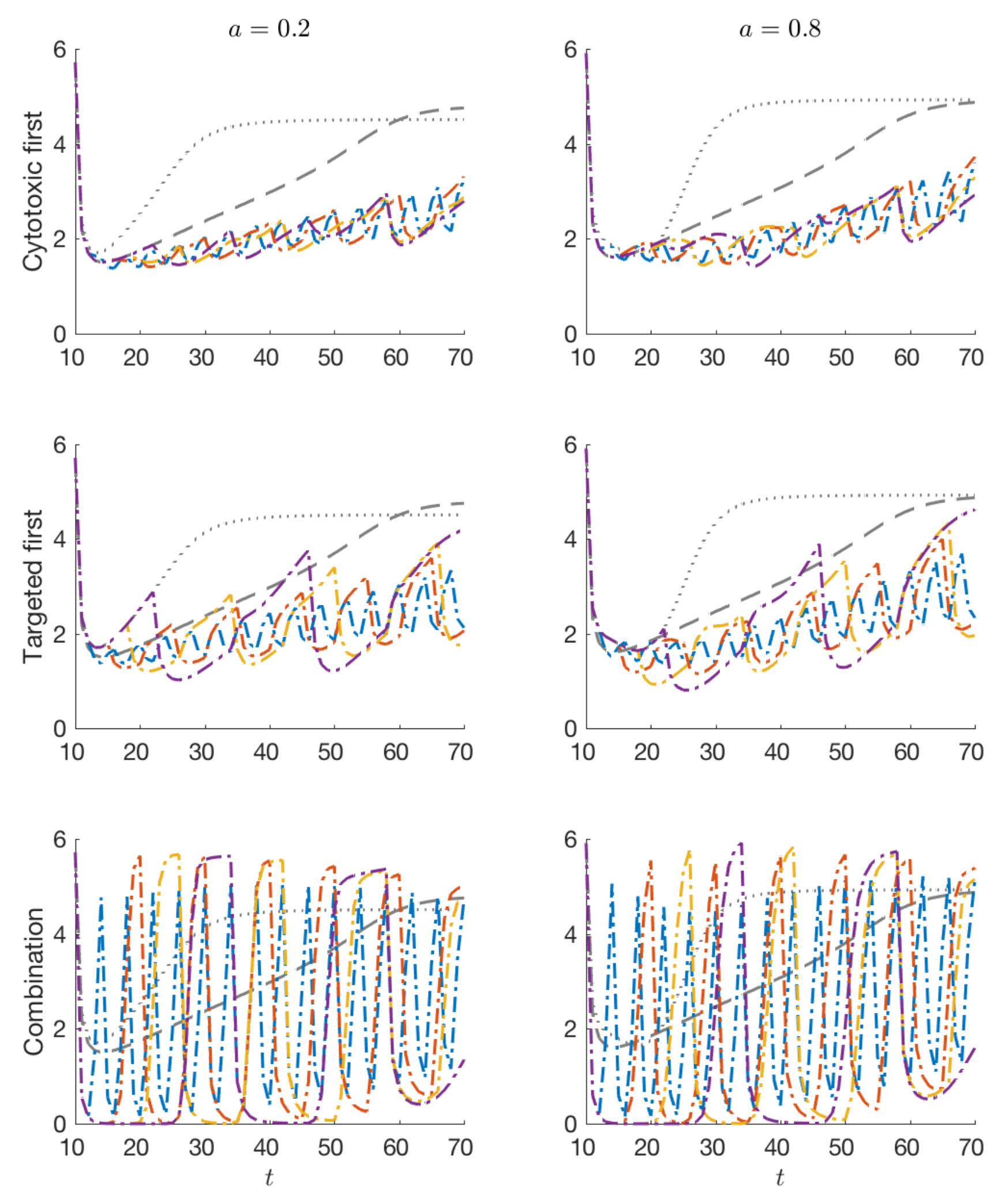} 
    }            \vspace{.1cm} 
    \centerline{  \rotatebox{90}{\hspace{2.9cm} \footnotesize $c_1+c_2 = 5$  }  \hspace{.2cm}      
       \includegraphics[width=7cm]{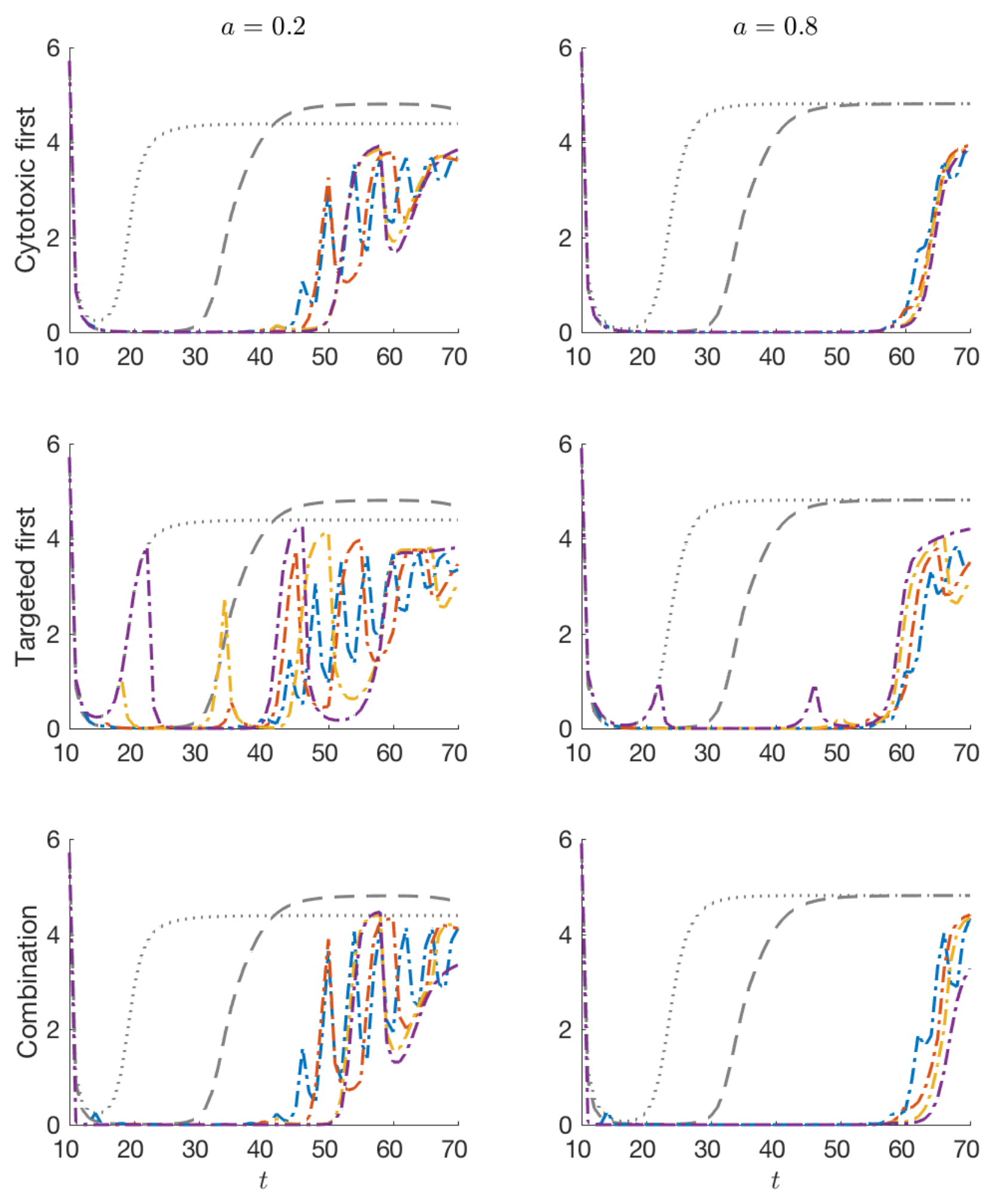} 
    }  
    \centerline{  \hspace{0.4cm}
       \includegraphics[width=6.5cm]{180514_nTotC_SwitchallP_ncase4_legend_.pdf} 
    }
   \caption{Number of cancer cells $\rho_c(t)$ 
   for different therapies when 
   the number of cells that are resistant to the targeted drug is
   larger than those that are resistant to the cytotoxic drug ($\rho_{c,R1} < \rho_{c,R2}$). 
   In contrast to Figure~\ref{fig:18}, initiating the alternating therapy with 
   the cytotoxic drug is more effective. 
     }
\label{fig:19} 
\end{figure}

Figure~\ref{fig:18} presents the results for the case
when the cytotoxic resistant cells have a higher ratio, 
that is, $\rho_{c,R1}(0) > \rho_{c,R2}(0)$.
Figure~\ref{fig:19} shows the opposite case.
We test the same therapies as before including 
the single drug, alternating therapies, 
and combination therapies.  We consider the moderate dosage
$c_1+c_2=3$ and the high dosage $c_1+c_2=5$.  The competition rates
are set as $a=0.2$ and $0.8$.  
Similar conclusions hold as in the symmetric pre-treatment
case.  With a relatively low dosage, 
the alternating schedule with a small period works remarkably better than 
the combination therapy, while with a relatively high dosage, 
the combination therapy can also suppress the tumor growth 
for a certain period of time. 
However, in the asymmetric pre-treatment case, 
the order of drugs becomes more important to
the therapy outcome.  First,
the outcome of single drug therapy is correlated with
the size of the preexisting resistance:
In Figure~\ref{fig:18}, 
the cytotoxic drug produces a worse outcome due to a larger pre-treatment
resistance population, while 
the targeted drug yields the early relapse portrayed in Figure~\ref{fig:19}.  
In addition, when $\rho_{c,R1}(0) > \rho_{c,R2}(0)$, initiating
an alternating schedule with a
targeted drug is more effective than initiating it with the cytotoxic drug, 
particularly for higher dosages.
Clearly, this is the outcome because the targeted drug reduces the
population of cells that are resistant to the cytotoxic drug.
In addition, for the highly competitive case, $a=0.8$, 
we observe that a single targeted drug therapy with dosage 
$c_1+c_2 = 3$ yields the minimal number of cancer cells up to $t \approx 40$ 
with a relatively low dosage.  This provides us with an opportunity to design
an effective adaptive therapy. 

On the other hand, when $\rho_{c,R1}(0) < \rho_{c,R2}(0)$, it is 
better to initiate the treatment with the cytotoxic drug. 
As expected, the results suggest that the pretreatment drug
resistance can be a critical factor in determining the coure of
therapy and its outcome.

\begin{figure}[!htb]
    \centerline{  \rotatebox{90}{\hspace{0.8cm} \footnotesize $\rho_{c,R1} < \rho_{c,R2}$ \hspace{1.2cm} $\rho_{c,R1} > \rho_{c,R2}$  }    
       \includegraphics[width=6.7cm]{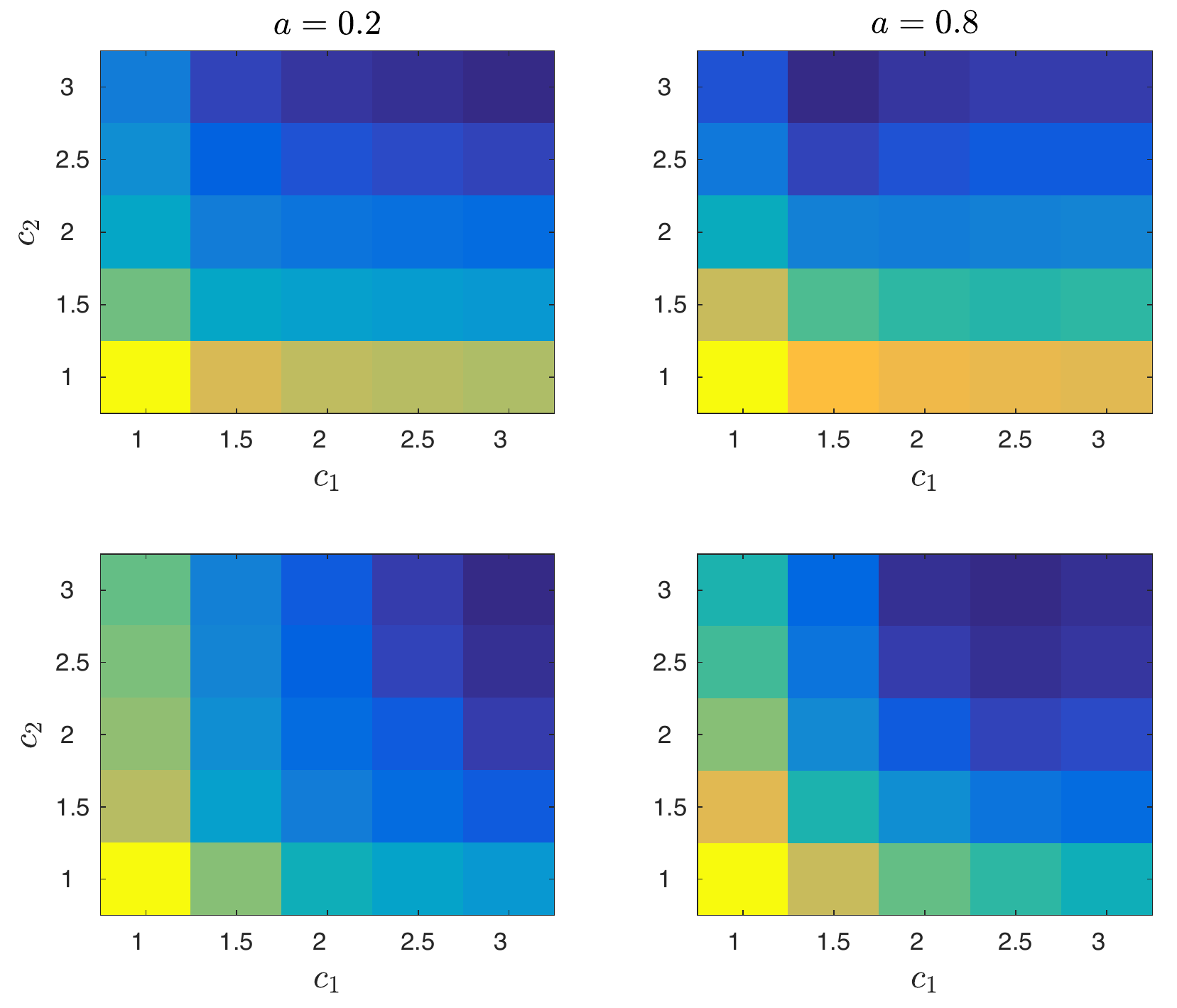} 
    }
   \caption{ Number of cancer cells at time $t=100$, $\rho_c(100)$,
     for different dosages of combination therapy involving cytotoxic
     and targeted drugs. The results are tested for 
     different sizes of pre-existing resistance either $\rho_{c,R1} < \rho_{c,R2}$ 
     or $\rho_{c,R1} > \rho_{c,R2}$. 
     As expected, initiating 
     the switching therapy with the drug that has a smaller resistant population 
     is more effective. When $a=0.8$, there exists an optimal dosage of
     the cytotoxic drug. 
     }
\label{fig:20} 
\end{figure}

In addition to drug scheduling, we also study the effect of 
dosages with respect to asymmetric 
preexisting resistance. In Figure~\ref{fig:20}, 
we present the number of cancer cells $\rho_c(t)$ at $t=100$ 
using a combination therapy with a cytotoxic drug dosage $c_1$ and 
a targeted drug dosage $c_2$. 
When $\rho_{c,R1} > \rho_{c,R2}$, 
we observe that a higher dosage of the targeted drug is more 
effective than increasing the dosage of the cytotoxic drug.
For instance, the dosage $(c_1,c_2) = (2,3)$ 
results in a smaller tumor than $(c_1,c_2) = (3,2)$.
In the opposite case, $\rho_{c,R1} < \rho_{c,R2}$, the result is reversed:
$c_1 < c_2$ is a more effective treatment.
When the competition is mild, $a=0.2$, 
increasing the dosages of both drugs constantly improves the outcome.
However in the highly competitive case, $a=0.8$, 
there exists an optimal dosage of the cytotoxic drug. 
For instance, when $\rho_{c,R1} > \rho_{c,R2}$, 
$(c_1,c_2)=(1.5,3)$ results with the minimum number of cancer cells. The
optimal dosage changes to $(c_1,c_2)=(2.5,3)$, with a slightly 
larger cytotoxic drug dosage when $\rho_{c,R1} < \rho_{c,R2}$. 

\section{Tumor growth model with cell competition} \label{sec:Num4} 

We extend the competition model by including a physical space variable 
$x \in [-1,\,1]^2 \subset \R^2$. The concentrations of 
healthy cells, $n_h(t,x,\theta)$, and cancer cells, $n_c(t,x,\theta)$, are governed 
by the following system, 
\begin{align}
   	 \partial_t n_h(t,x,\theta) &= G_H n_h	+  \nu_n \Delta_x n_h + \nu_p \nabla_x \cdot ( n_h \nabla_x p_h ), \\ \nonumber 
   	 \partial_t n_c(t,x,\theta) &= G_C n_c	+ \nu_n \Delta_x n_c + \nu_p \nabla_x \cdot ( n_c \nabla_x p_c ). 
  \label{eq:PDEsp} 
\end{align} 
Here, $p_h(t,x) = (\rho_h/\rho_{h,0})^k$ and 
$p_c(t,x) = (\rho_c/\rho_{c,0})^k$, 
are the cell pressures for the healthy cells and the cancer cells,
respectively.
The normalizing constants
are taken as the maximum cell capacity 
$\rho_{h,0}=3$ and $\rho_{c,0}=6$. 
The growth terms, 
$G_H$ and $G_C$, are taken as 
in Eqs.~\eqref{eq:GvrnH0} and \eqref{eq:GvrnC0}, 
and $\nu_n$ and $\nu_p$ are constants describing 
cell motility. 
The spatial competition model follows the tumor growth model 
developed in \cite{Cho2017a}, 
and the cell motility parameters are taken as 
$\nu_h = \nu_c = 10^{-6}$, $\nu_p = 10^{-5}$, and $k=6$
\citep{Bray}.

\begin{figure}[!thb]
    \centerline{ \rotatebox{90}{\hspace{0.7cm}\scriptsize $a=0.2$  }
    \rotatebox{90}{\hspace{0.5cm}\scriptsize Cytotoxic  }
    \includegraphics[width=7.5cm]{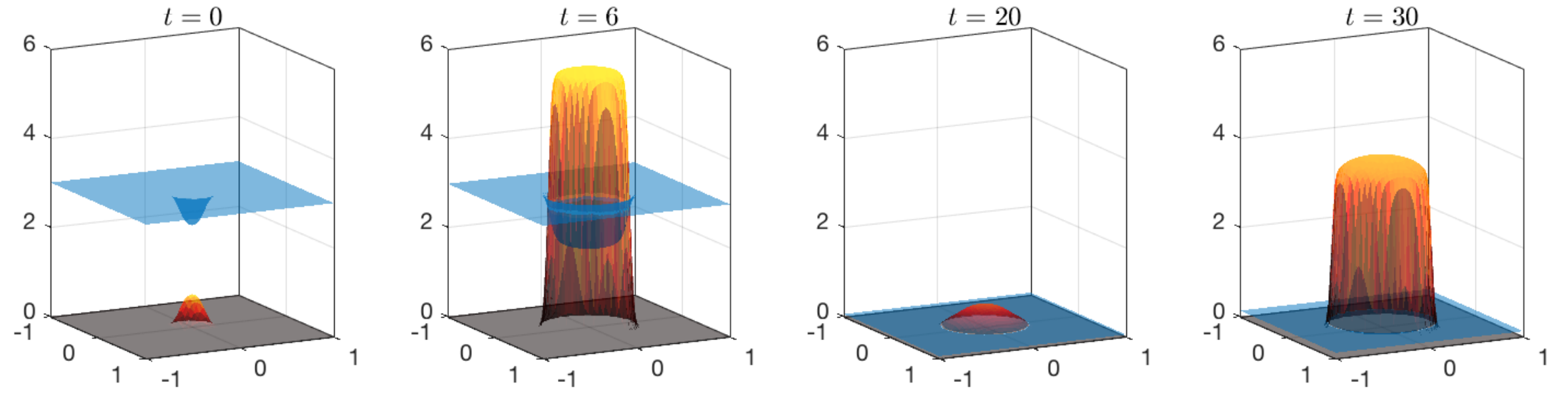} } 
    \centerline{ \rotatebox{90}{\hspace{0.7cm}\scriptsize $a=0.8$  }
    \rotatebox{90}{\hspace{0.6cm}\scriptsize Targeted  }
    \includegraphics[width=7.5cm]{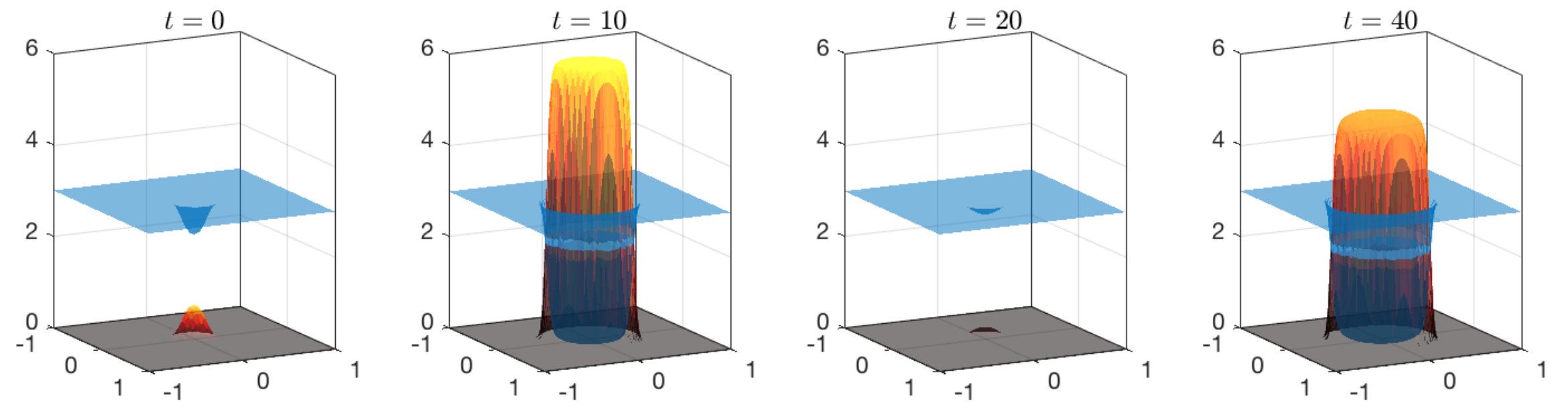} } 
   \caption{ The evolution of cancer cells $\rho_c(t,x,y)$ and 
   healthy cells $\rho_h(t,x,y)$ under single-drug therapies 
   initiated with a small cancer population in the center of the domain (left, $t=0$). 
   The cancer cells grow on top of the healthy tissue 
   when $a=0.2$ (top, $t=6$). 
   When $a=0.8$ (bottom, $t=10$), 
   the cancer cells aggressively eliminate 
   the healthy cells while expanding.  The treatment starts at 
   $t=6$ (top) and $t=10$ (bottom).
   The cancer cells relapse quickly when 
   using a single drug therapy due to preexisting resistance. 
   }  
\label{fig:22}
\end{figure}

We consider three spatially heterogeneous 
drug distributions to 
examine the therapies $c_1(t,x)$ and $c_2(t,x)$:
\begin{enumerate}
\item[i.] A constant dosage, 
$$c_i(t,x) = \bar{c}_i(t).$$
\item[ii.] A diffusive case, where the drug diffuses from the right edge $x_1=1$
\citep{Mumenthaler2015},
$$c_i(t,x) = \bar{c}_i(t) \left[
(e^{\lambda (x_1+1)/2} + e^{-\lambda (x_1+1)/2}) 
/( e^{\lambda} + e^{-\lambda}) \right],$$  with
$\lambda = \sqrt{2}$.
\item[iii.] A highly heterogeneous case \citep{Peng2016}, 
\begin{align*}
c_i(t,x) &= \bar{c}_i(t) \Big[ 2 + 0.25 \sin( 2\pi\, \|\, (x_1+1 , x_2+1)\, \|_2 ) 
\Big. \\  
&+ \Big. 0.5 \sin( 4\pi\, \|\,(1-x_1 , x_2+1)\, \|_2 ) \Big]/ 2.75. 
\end{align*}
\end{enumerate}
We assume a similar dependence of
the resources on the space variable,
$\eta_h(x)$ and $\eta_c(x)$, with all three cases considered.
We choose 
the initial condition as a small concentration 
of cancer cells embedded 
in the center of a healthy tissue
(see Figure \ref{fig:22}, $t=0$). 
The following results are presented by plotting
the total number of healthy cells, $\rho_h(t,x,y) = \int n_h(t,x,y, \theta) d\theta$,
and cancer cells, $\rho_c(t,x,y) = \int n_c(t,x,y, \theta) d\theta$.

Figure \ref{fig:22} corresponds to the case of a constant dosage.
The initial tumor is located at the center of the domain.
The first row shows the mildly competitive case, $a=0.2$,
and treatment with the cytotoxic drug. The second row shows the highly competitive case, 
$a=0.8$, and treatment with the targeted drug.
The treatments are 
initiated at $t_c = 6$ and $t_c = 10$, respectively, 
with dosages $\bar{c}_1 = \bar{c}_2 = 5$. 
We observe that 
the cytotoxic therapy eliminates the healthy tissue in addition to 
the cancer cells, unlike the targeted drug. 
In addition, when $a=0.2$, the cancer cells grow on top of 
the healthy tissue, while in 
the highly competitive case, $a=0.8$, the cancer cells replace
the healthy tissue while expanding. 
In both cases, a tumor treated with a single drug therapy
quickly relapses 
due to the preexisting resistant cells. 
The spatial simulations 
with constant dosages are consistent with the results 
of the non-spatial model in the previous sections.

\begin{figure}[!htb]
    \centerline{ \rotatebox{90}{\hspace{0.4cm}\scriptsize Cytotoxic }
       \includegraphics[width=7cm]{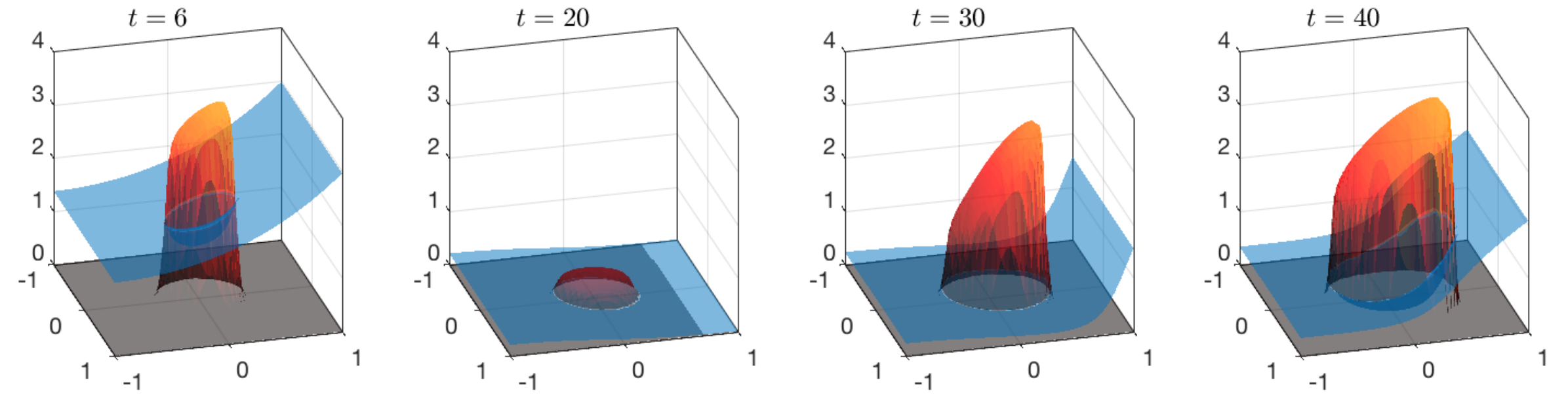} 
    }
    \centerline{ \rotatebox{90}{\hspace{0.3cm}\scriptsize Alt. \hspace{0.0cm} $a=0.2$ }
       \includegraphics[width=7cm]{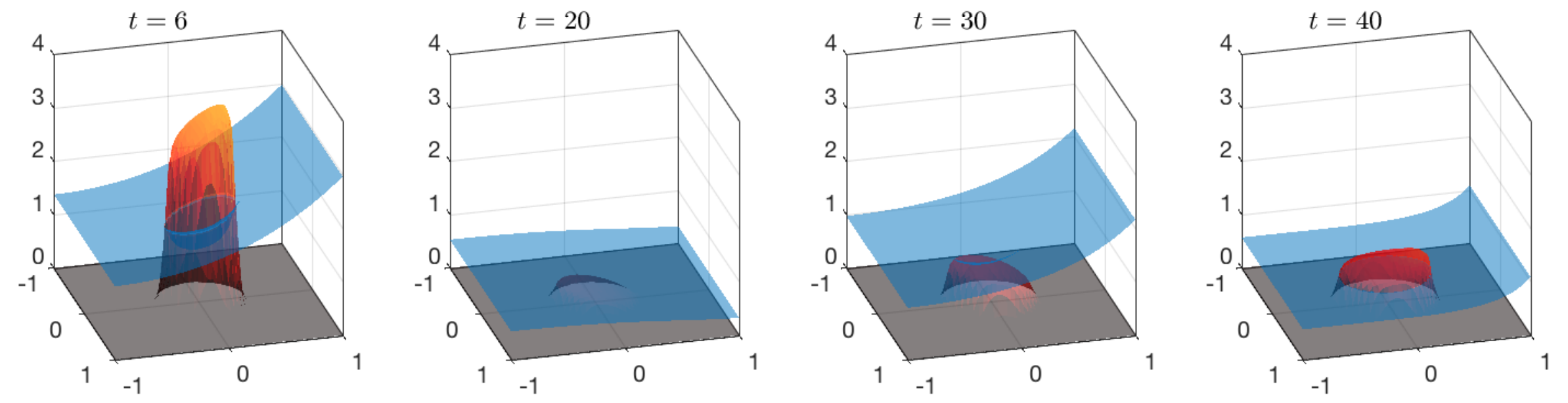} 
    }
    \centerline{ \rotatebox{90}{\hspace{0.4cm}\scriptsize Targeted }
       \includegraphics[width=7cm]{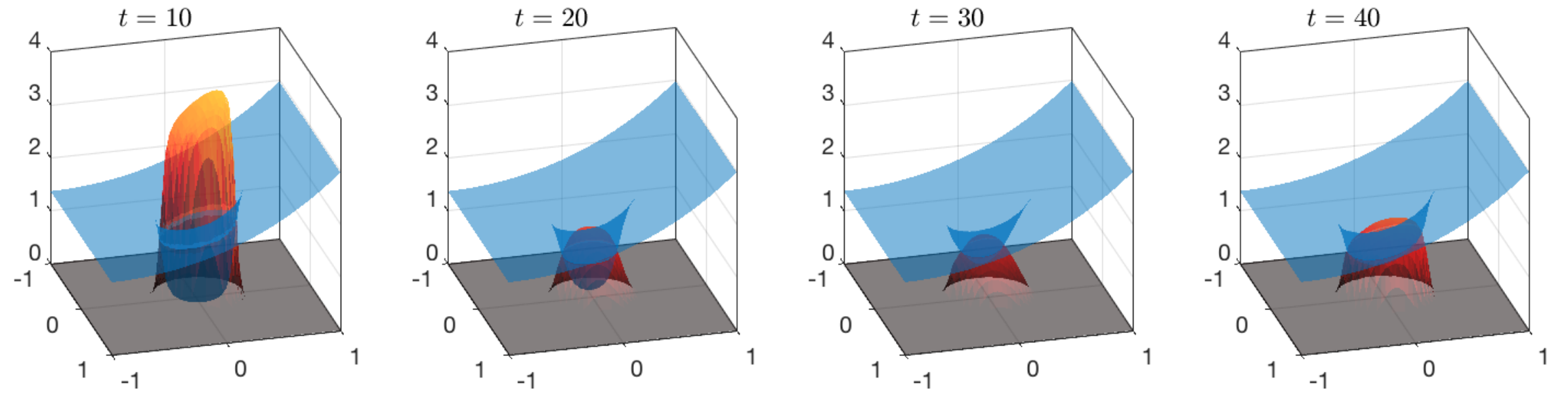} 
    }
    \centerline{ \rotatebox{90}{\hspace{0.3cm}\scriptsize Alt. \hspace{0.0cm} $a=0.8$ }
       \includegraphics[width=7cm]{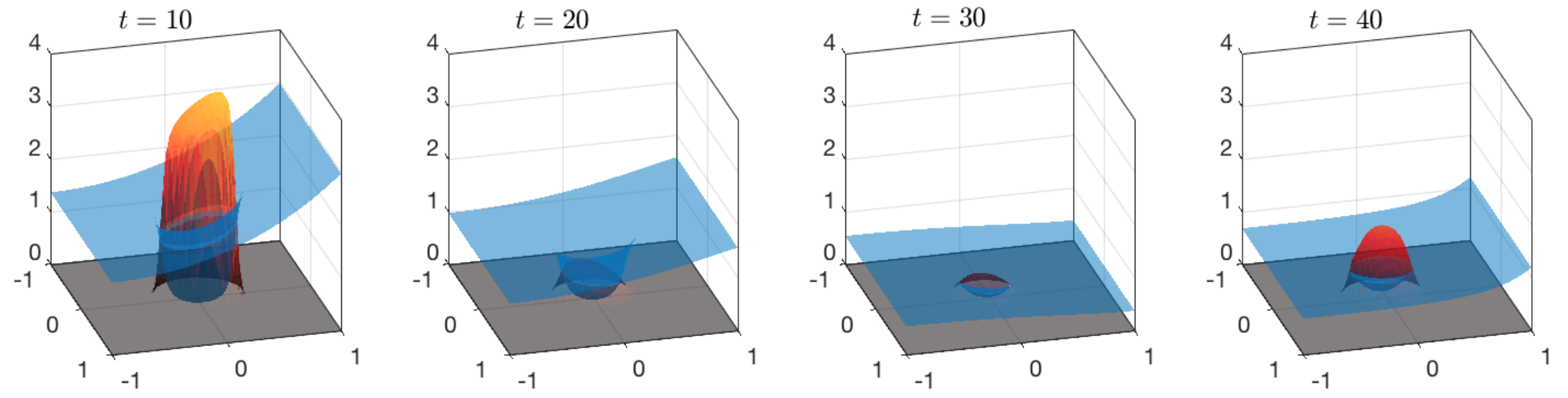} 
    }
   \caption{ The evolution of cancer cells $\rho_c(t,x,y)$ and 
   healthy cells $\rho_h(t,x,y)$ under single-drug and alternating (Alt.) 
   therapies diffused from the right boundary, $x=1$. 
   When $a=0.2$, the alternating therapy is remarkably effective compared with
   a single cytotoxic drug. When $a=0.8$, a single targeted drug therapy 
   is also effective, although the tumor size is slightly larger
   compared with the tumor size with the alternating therapy.  }  
\label{fig:23}
\end{figure}

We now test combination therapies when the drug 
distribution is spatially heterogeneous. 
Figure \ref{fig:23} compares single drug therapies and 
alternating therapies with period $t_p=2$, 
when the resource and drugs are diffused 
from the right boundary, $x_1 = 1$. 
The treatment dosages are taken as 
$\bar{c}_1 = 7$ and $\bar{c}_2 = 4$.  
We remark that prior to the treatment, the tumor grows faster
closer to the right boundary 
where the concentration of resources is high. 
Using a single drug therapy, the tumor relapses before $t=40$, 
particularly when  $a=0.2$ using the cytotoxic drug. 
When $a=0.8$, we verify the effectiveness of 
the targeted drug, for which we observe that 
the size of the tumor at $t=40$ is smaller 
compared with the tumor at the same time using the cytotoxic drug, despite the 
lower drug dosage. 
The alternating therapies are
more effective compared with the single-drug therapies 
in all competition environments,
although the difference is smaller
in the highly competitive case $a=0.8$, 
since the single targeted drug therapy is effective as well. 

\begin{figure}[!htb]
    \centerline{ \rotatebox{90}{\hspace{0.4cm}\scriptsize   Cytotoxic }
       \includegraphics[width=7cm]{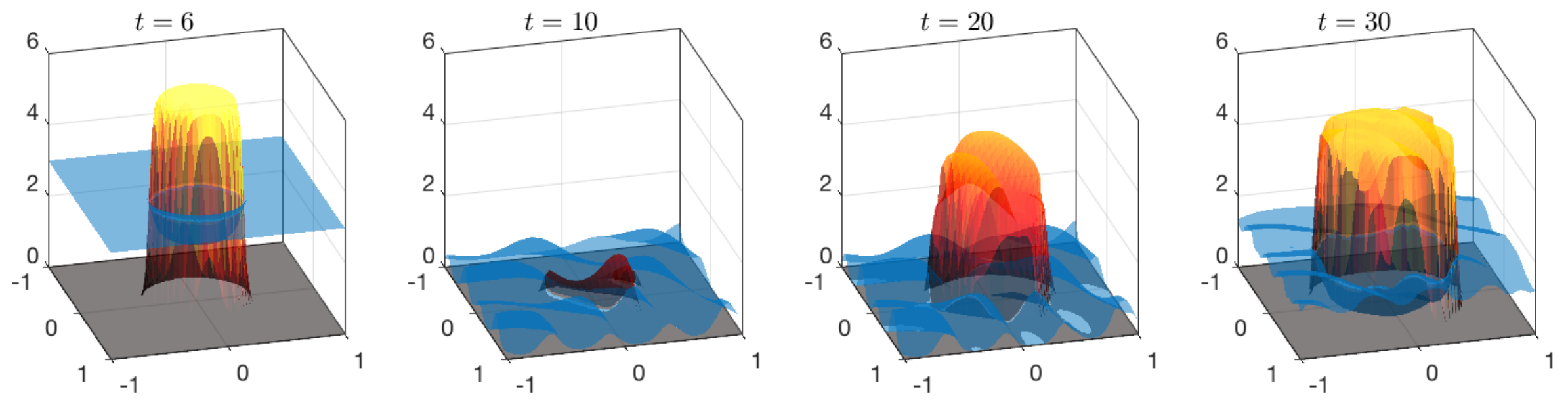} 
    }
    \centerline{ 
    \rotatebox{90}{\hspace{0.4cm}\scriptsize $a=0.2$  }
    \rotatebox{90}{\hspace{0.6cm}\scriptsize Alt. }
       \includegraphics[width=7cm]{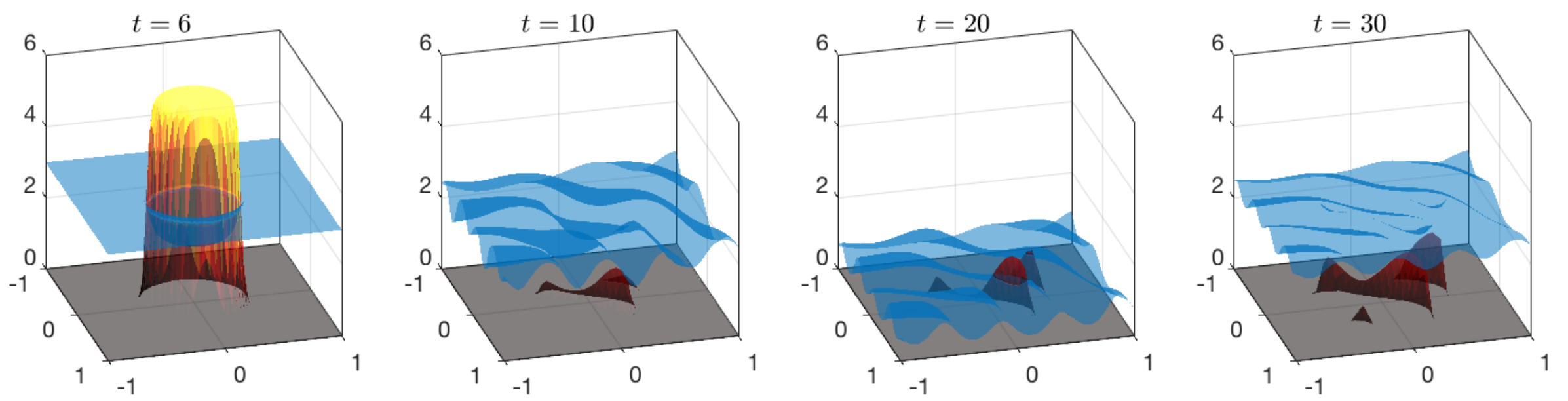} 
    }
    \centerline{ \rotatebox{90}{\hspace{0.4cm}\scriptsize Comb. }    
       \includegraphics[width=7cm]{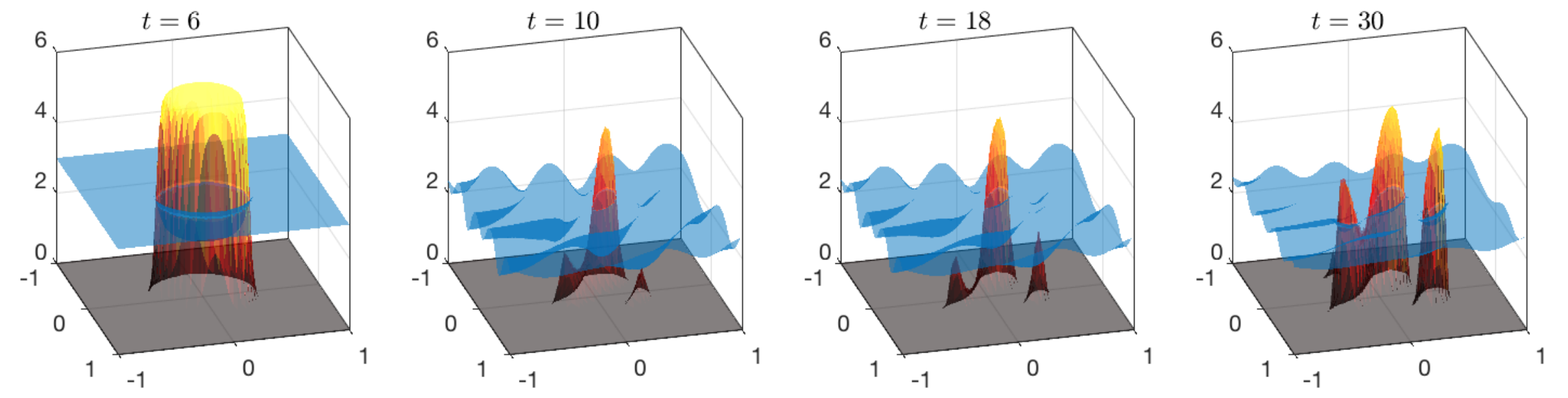} 
    }   
    \centerline{ \rotatebox{90}{\hspace{0.4cm}\scriptsize Targeted }
       \includegraphics[width=7cm]{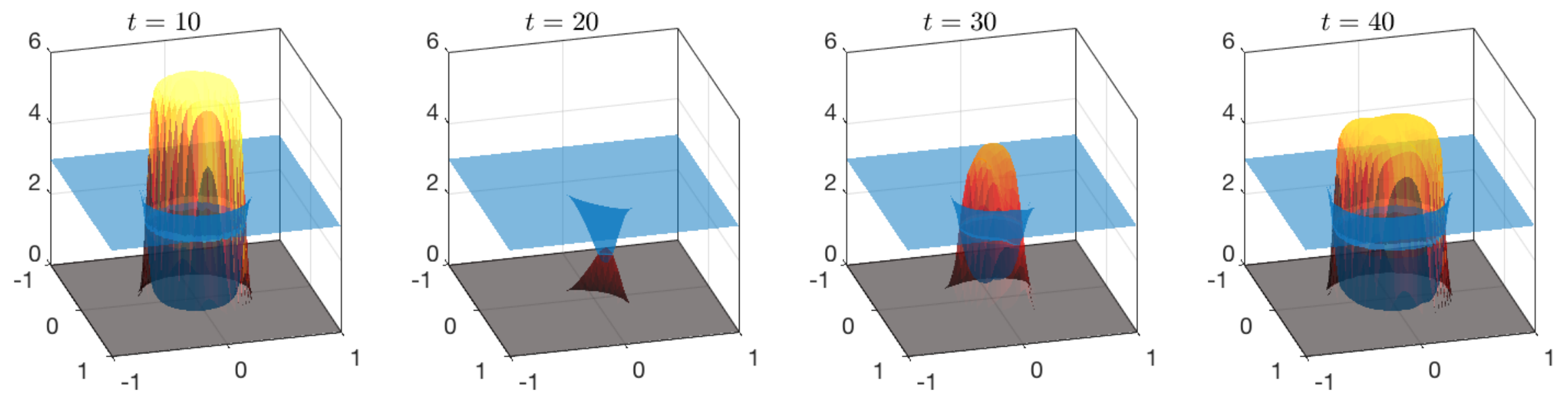} 
    }
    \centerline{ \rotatebox{90}{\hspace{0.4cm}\scriptsize  $a=0.8$ }
    \rotatebox{90}{\hspace{0.6cm}\scriptsize Alt. }
       \includegraphics[width=7cm]{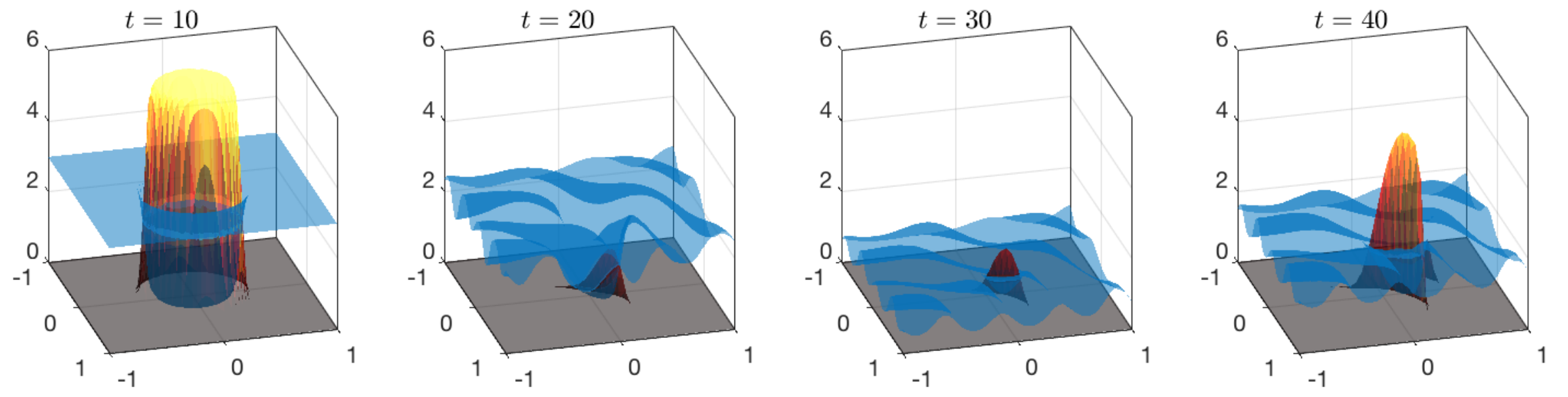} 
    }
    \centerline{ \rotatebox{90}{\hspace{0.4cm}\scriptsize Comb. }
       \includegraphics[width=7cm]{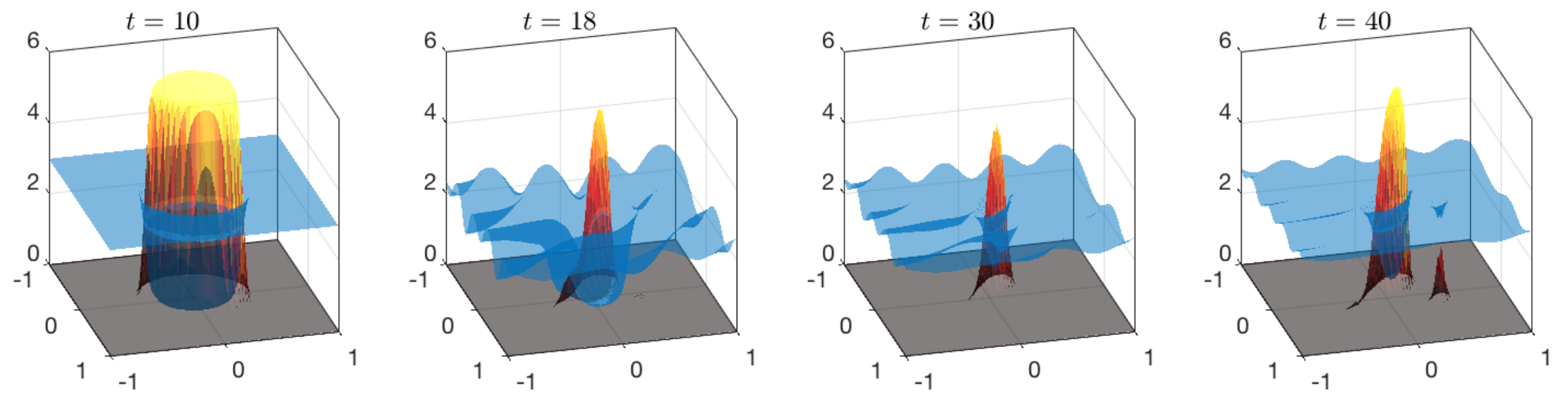} 
    }       
   \caption{ The evolution of cancer cells $\rho_c(t,x,y)$ and 
   healthy cells $\rho_h(t,x,y)$ under different therapies 
   with an irregular drug distribution. 
   Local peaks of cancer cells can be observed that eventually 
   grows into a tumor with a rough surface. 
   The single-drug therapies result in strong relapses 
   compared with the outcomes of alternating (Alt.) 
   and combination (Comb.) therapies. 
   Moreover, the alternating therapy is more effective than 
   the on-off combination therapy particularly 
   when $a=0.2$. 
 }  
\label{fig:24}
\end{figure}

Finally, different therapies including 
the on-off combination therapies are compared in Figure \ref{fig:24}. 
We set the dosages as $\bar{c}_1 = \bar{c}_2 = 5$, which are 
sufficiently high so that the on-off combination therapies 
are effective as much as the alternating therapies. 
The drug distribution is heterogeneous.
As expected, the single drug therapies result with 
strong relapses 
compared with the alternating and combination therapies. 
In addition, we observe emerging local peaks of cancer cells 
when using combination therapies during the off periods. This is
particularly worse than the outcome of alternating therapies when
$a=0.2$.  In case of $a=0.8$, although the alternating therapy
is more effective in suppressing the tumor throughout the treatment 
than the combination therapy, 
the sizes of the relapsed tumors at $t=40$ are similar.

\section{Conclusion} \label{sec:conclusion}

In this work we develop a
competition model of healthy and cancer cells 
that takes into account resistance to cytotoxic and targeted 
drugs.
We  study the dynamics of resistance 
to the drugs and
observe the emergence of populations with distinct levels of resistance
depending on the therapy. 
Primarily, we classify the cell competition scenarios 
as either mild, where distinct cell types can coexist, 
or aggressive, where cancer cells dominate
by actively eliminating the healthy cells.
The threshold of the competition rate that distinguishes the two scenarios 
is related to the over-proliferation of the 
cancer cells over the healthy tissue. It also depends on the 
drug dosages. 
In addition, the analysis shows that 
targeted therapies have a greater potential of being effective
when the cells are highly competitive.

Various drug treatments are tested in the two competition scenarios, 
and we observe that the treatment outcomes are distinctive. 
Although the targeted drug is more effective in the highly competitive case, 
using a single drug therapy, either cytotoxic or targeted, 
results with an eventual relapse 
due to the preexisting resistance, regardless 
of the strength of the competition.
However, treatments that include both drugs show better outcomes 
in terms of the relapse time and the tumor size. 
Considering the drug switching therapy, 
an optimal switching time that 
minimizes the overall number of cancer cells 
exists when the competition is mild. In the
highly competitive case, the targeted drug therapy alone 
is often effective enough. 
We also compare different continuum models that 
either allow for intermediate resistance states
or are close to a two-state model with cells that are either
fully sensitive or fully resistant to the drugs.
Although the overall advantage of the switching therapy 
over single drug therapy 
in different competition environment holds, 
the linear model is shown to be more sensitive to 
the switching time that often yields a worse outcome 
compared with a single targeted drug therapy. Thus, when 
the population is highly competitive and 
the tumor proliferation and the drug uptake linearly depend
on the resistance trait, 
the drug switching time should be more carefully determined.
Alternating treatments with different periods are 
shown to be effective in suppressing 
the cancer population during the entire treatment period 
compared to the other therapies. 
This particularly holds with small periods.
Finally, we investigate a spatially heterogeneous tumor growth model, 
and verify that the same conclusions hold.

As future work, 
we propose to incorporate experimental results 
considering 
combination of chemotherapy and targeted therapies
\citep{Dorris2017,Ribeiro2012} 
and develop optimal strategies 
using optimal control theory 
for stabilizing the cancer population and/or
minimizing the tumor size during the treatment period 
\citep{Carrere2017, Jonsson2017}. 
Adaptive therapy 
is another interesting topic 
that aims at controlling the tumor 
by maintaining sensitive cells in order to suppress the 
resistant cancer cells \citep{Gatenby2009,Bacevic2017}. 
Finally, the computational cost of simulating
three-dimensional tumor growth models with 
multi-dimensional resistance traits is prohibitively
expensive due to the high dimensionality. This requires 
developing an efficient numerical method 
that balances computational cost and accuracy \citep{Grasedyck2013,Cho2016}.

\section*{Acknowledgments}
The work of DL was supported in part by the National Science
Foundation under Grant Number DMS-1713109 
and by the Jayne Koskinas Ted Giovanis Foundation.

\bibliographystyle{elsarticle-harv}
\bibliography{Tumor}
\biboptions{sort&compress}

\end{document}